\documentclass[12pt,preprint]{aastex}

\newcommand{\kms}{km~s$^{-1}\:$}
\newcommand{\Ha}{H$\,\alpha\:$}
\newcommand{\HI}{H\,I\,\,}
\newcommand{\Hp}{H$^+\:$}
\newcommand{\NHI}{$N_{\mathrm{H\,I}}\,$}
\newcommand{\cmm}{cm$^{-2}\:$}
\newcommand{\cmmm}{cm$^{-3}\:$}
\newcommand{\be}{\begin{equation}}
\newcommand{\ee}{\end{equation}}

\slugcomment{Accepted for publication in the Astrophysical
Journal}

\shorttitle{Ophiuchus Superbubble}

\shortauthors{Pidopryhora et~al.}

\begin{document}

\title{The Ophiuchus Superbubble: A Gigantic Eruption from the Inner Disk of the Milky Way}
\author{Yurii Pidopryhora\altaffilmark{1,2}, Felix J. Lockman\altaffilmark{1}, Joseph C. Shields\altaffilmark{2}}

\email{ypidopry@nrao.edu, jlockman@nrao.edu,
shields@phy.ohiou.edu}

\altaffiltext{1}{National Radio Astronomy Observatory, PO Box 2,
Green Bank, WV 24944}

\altaffiltext{2}{Astrophysical Institute, Department of Physics \&
Astronomy, Clippinger Laboratories, Ohio University, Athens, OH
45701}

\begin{abstract}

While studying extraplanar neutral hydrogen in the disk-halo
transition of the inner Galaxy we have discovered what appears to
be a huge superbubble centered around $\ell \approx 30\arcdeg$,
whose top extends to latitudes $>25\arcdeg$ at a distance of about
7~kpc.  It is detected in both \HI and \Ha. Using the Green Bank
Telescope of the NRAO, we have measured more than 220,000 \HI
spectra at $9\arcmin$ angular resolution in and around this
structure. The total \HI mass in the system is $\approx
10^6$~$M_{\sun}$ and it has an equal mass in \Hp.  The Plume of
\HI capping its top is $1.2 \times 0.6$~kpc in $\ell$ and $b$ and
contains $3 \times 10^4$~$M_{\sun}$ of \HI. Despite its location,
(the main section is 3.4~kpc above the Galactic plane) the
kinematics of the Plume appears to be dominated by Galactic
rotation, but with a lag of 27~\kms from corotation.  At the base
of this structure there are ``whiskers'' of \HI several hundreds
of pc wide, reaching more than 1~kpc into the halo; they have a
vertical density structure suggesting that they are the bubble
walls and have been created by sideways rather than upwards
motion.  They resemble the vertical dust lanes seen in NGC 891.
From a Kompaneets model of an expanding bubble, we
estimate that the age of this system is $\approx 30$~Myr and its
total energy content $\sim 10^{53}$~ergs.  It may just now be at
the stage where its expansion has ceased and the shell is
beginning to undergo significant instabilities.  This system
offers an unprecedented opportunity to study a number of important
phenomena at close range, including superbubble evolution,
turbulence in an \HI shell, and the magnitude of the ionizing flux above the
Galactic disk.

\end{abstract}


\section{INTRODUCTION}

A key factor in Galactic evolution is the very complex interaction
between stellar evolution and the interstellar medium.  Stellar
winds and supernovae not only redistribute hydrogen and heavy
elements, but in sufficient quantity, can totally shut down star
formation by moving gas to the intergalactic medium.  In addition,
circulation of gas from the disk into the halo, or accretion of
low-metallicity gas, will alter the chemical evolution of the
Galaxy (\citet{GalWinds} and references therein; \citet{Sancisi,
Tripp}).

The Milky Way offers a unique yet often confused perspective on
these processes.  Neutral, warm, and hot gas is observed far from
the Galactic plane, and `superbubbles' are sometimes seen
surrounding sites of recent star formation, though the connection
between the different phases, and indeed the differentiation
between local, Galactic halo, and intergalactic phenomena is not
always clear.

A search  for shells and shell-like structures in the Milky Way
was completed by \citet{Heiles79, Heiles84}, complemented by
\citet{Hu81} at high Galactic latitudes and by
\citet{McClure-Griffiths} in the southern hemisphere. Recently an
automated search for shells in the Leiden-Dwingeloo survey
\citep{LD} was carried out by \citet{Ehlerova}, who discovered
nearly 300 structures, several of which were identified with
objects from Heiles' and Hu's catalogs. Shells and supershells are
also observed in nearby galaxies including M31, the LMC and the
SMC
\citep{Brinks86,Kim,Staveley97,Stanimirovic99,WalterBrinks,Howk}.
 \citet{Heiles84} and
\citet{McClure-Griffiths} thoroughly analyze a few specifically
chosen
 shells from their surveys \citep[see also][]{McClure2000}. Recent
 examples of studies of individual shells are: Aquilla Supershell:
 \citet{Maciejewski96}; Scutum Supershell: \citet{Callaway2000} (in \HI,
 \Ha, infrared, X-ray and UV), \citet{Savage2001} (UV-absorption).

The simplest spherically-symmetric self-similar theory of the
superbubbles was developed by \citet{PikelnerShcheglov69},
\citet{SpherBubble} and \citet{Weaver}, among others.
\citet{KooMcKee1,KooMcKee2} gave the problem of
spherically-symmetric bubble a more general analytical treatment,
and \citet{OeyClarke} used the self-similar approach to derive the
size distribution of shells in a galaxy; their results are in good
agreement with observations of nearby galaxies. Based on an early
approach of \citet{Kompan60}, a 2D semi-analytical model of a
bubble in a stratified medium was developed by \citet{MacLow}.
\citet{TomisakaIkeuchi86,MacLow89,T-TBR,Tenorio90,Igumen90} gave
the 2D problem a numerical treatment. \citet{MacLow89} found the
results of approximate Kompaneets-model calculations to be in
excellent agreement with their exact numerical solutions. A
detailed review of early theoretical and observational work on
superbubbles was given by \citet{TenorioTagleARAA88}. A detailed
critical analysis of the Kompaneets approximation was given by
\citet{KooMcKee0}, who showed how this method can be improved. The
complete range of approximate superbubble models was reviewed by
\citet{Bisnovatyj}.

A number of 3D numerical treatments of the superbubble problem
have now been done: \citet{Tomisaka98} studied how a Galactic
magnetic field changes the traditional 2D numerical results;
\citet{Korpi99} simulated superbubbles with a non-ideal MHD model;
\citet{AvillezBerry2001} carried out a high-resolution
hydrodynamical study and their approach was expanded by
\citet{Breitschwerdt06} to model the Local and Loop~I bubbles.

In a recent discussion \citet{Oey2004} finds the conventional
superbubble paradigm to be consistent with the observational data.
A few problems exist, the most serious of which is the
``energy-deficit problem'' \citep{OeyGarcia,Cooper04}, but there
is nothing which suggests that the basic theory is in error.

New studies in directions where one might have some hope of
untangling the different processes and understanding their
relationships may yield interesting results. To this end we have
been measuring the disk-halo transition of the Milky Way in the
21~cm line of \HI, using the Green Bank Telescope to map neutral
gas above the Galactic plane in the inner Galaxy. The power of
this telescope is such that significant new insights into Galactic
\HI can be obtained with integration times as short as two seconds
per spectrum. We have focussed our first efforts on regions near
the tangent points in the first longitude quadrant of the inner
Galaxy, where Galactic rotation is projected entirely along the
line of sight and the distance to the highest velocity features
can be estimated with a reasonable degree of accuracy. This paper
reports the discovery of a very large, coherent \HI structure
which extends $>3$~kpc into the Galactic halo near the tangent
point at $\ell = 30\arcdeg$, and which is likely to be a
relatively nearby example of the superbubbles playing a critical
role in galaxy evolution. It lies predominantly in the Ophiuchus
constellation, so we refer to it as the Ophiuchus superbubble.

In section 2 of this paper we describe the \HI observations and in
section 3 the corresponding data reduction; section 4 summarizes
the observed properties of the superbubble in both \HI and \Ha. In
section 5 we discuss the distance to the system and derive some of
its physical properties. Section 6 tests the validity of the
superbubble hypothesis against a simple analytical model, which
also provides estimates of the age and energetics of the system,
and in section 7 possible sources of the superbubble's origin and
ionization are considered. Section 8 concludes the paper with a
general discussion of the results.

\section{OBSERVATIONS}

The observations were made with the National Radio Astronomy
Observatory's 100~m diameter Robert C. Byrd Green Bank Telescope
(GBT)\footnotemark.  The angular resolution of the telescope at
21~cm wavelength is $9\farcm 1$ (FWHM). The receiver was dual
circularly polarized and had a system temperature at zenith of 18
K. Spectra were measured using the Spectral Processor, a $2 \times
1024$ channel FFT spectrometer  operated at a bandwidth of 5~MHz
to give a channel spacing of 1.03~\kms and an effective velocity
resolution of 1.25~\kms.  Spectra were obtained by frequency
switching `in-band' for a total velocity coverage of about
500~\kms centered around $+50$~\kms (LSR).  With this arrangement
the rms noise in an individual channel for measurements made at
elevations $\gtrsim 20\arcdeg$ is $\approx 0.36 $ $t^{-1/2}$ K,
where $t$ is the integration time in seconds.

\footnotetext{The NRAO is operated by Associated Universities, Inc.,
 under a cooperative agreement with the National Science Foundation.}

The region of our investigation was mapped in segments, each
typically $2\arcdeg \times 2\arcdeg$ in Galactic longitude $\ell$
and latitude $b$, at a spacing of $3\arcmin$ in both coordinates.
This is slightly finer than the Nyquist sampling for the angular
resolution of the GBT. Data were taken while the telescope moved
in Galactic longitude at a rate which gave integration times of 2
seconds at each pointing position. Areas of special interest were
reobserved for an additional  5 seconds at each position, so the final
maps are a mix of data whose noise levels vary by factors of a
few when all effects are taken into account.  In all, more than
220,000 independent \HI spectra were obtained.  We report here on
just a fraction of the data in these measurements: the emission
from a large structure near the tangent point at $\ell \sim
30\arcdeg$.

\section{DATA REDUCTION}

Spectra were reduced and maps made using the aips++ software package
and its set of GBT functions {\it gbdish}.  Calibration was
accomplished through laboratory measurements of the receiver's noise
diodes with checks against the standard regions S6 and S8 \citep{Williams}.
In this experiment noise, and not calibration uncertainty,
dominates the error budget.  Instrumental baselines were removed from
the individual spectra using 2nd or 3rd-order polynomials fit to
emission-free velocities.  The spectra were assembled into a data cube
in aips++ using gridding functions which produced little loss of
angular resolution.  However, the on-the-fly scanning
 reduces the effective angular resolution in the scanning
direction with the result that the final data have an effective
angular resolution (FWHM) in $\ell$ and $b$ of $9\farcm 8 \times
9\farcm 3$.

In addition to the new observations, we have used a small amount of
GBT archival data taken during a survey of the lower Galactic halo
with an identical instrumental configuration \citep{FJL02}.

\subsection{Correction for Stray Radiation}

The main problem encountered during data reduction was a
contamination of some of the data by stray 21~cm radiation.  At
certain sidereal times a forward spillover lobe of the GBT admits
some Galactic disk emission into spectra taken many degrees away
from the plane  (see \citet{FJLstray}).  The stray component
produces broad, weak, variable emission on which the real features
lie.  As a first-order correction for this effect, we renormalized
all the contaminated spectra to the Leiden-Dwingeloo (hereafter
LD) survey of \citet{LD} in the manner described in \cite{FJL86}
and \citet{FJL02stray}. The GBT data are convolved to the angular
resolution of the LD survey, and any difference between the
convolved GBT spectra and that of LD is assumed to be stray
radiation in the GBT data, which is then removed from individual
GBT spectra. Figure~1 shows a GBT spectrum (at $9\arcmin$
resolution) before and after this correction, and the LD spectrum
in the same direction.  The GBT spectrum has a line from a compact
\HI cloud which is not detectable in the larger ($36\arcmin$) LD
beam.

The stray radiation correction procedure sometimes created long,
narrow horizontal and vertical artifacts at the boundaries of the
survey sections. These are quite noticeable in some of the figures
but are essentially small zero-offsets with negligible effect on
the analysis.  Work is underway to better estimate
the stray radiation correction and remove these artifacts.
The \HI features we discuss here are discrete in space and
velocity and can easily be distinguished from the broad stray \HI
component. All measurements reported in this paper have been
checked for consistency with the uncorrected data.

\subsection{Error Estimates}

The rms noise of spectra in the final data cube is typically 0.25~K
for the 2~sec. survey regions, and as low as 0.15~K in the deeper
areas that have 7~sec. integration times. The error introduced into
the total column density, \NHI, by baseline uncertainties is
comparable with the error from noise. The stray radiation correction
also introduces an error in column density measurements which is
difficult to quantify. In this paper we focus on \HI emission features
which are discrete in position and velocity, so the uncertainty in
\NHI introduced by the stray radiation correction should be comparable
to that of noise and the instrumental baseline.  In all, the error in
\NHI for the features discussed here is $\approx 3
\times 10^{18}$ cm$^{-2}$ ($1\sigma$) in a single pixel,
 and about half of this is
uncorrelated from pixel to pixel.

\section{THE OPHIUCHUS SUPERBUBBLE: OBSERVED PROPERTIES}

\subsection{A Neutral Hydrogen Plume}

Figure~2 is an \HI column density map covering more than 300
square degrees derived from GBT spectra integrated over $60 \leq
V_{\rm LSR}\leq 100$~\kms. This range of velocities corresponds to
gas rotating with the Galaxy near the tangent points at these
longitudes, implying a distance from the Sun $r = R_0 \sin \ell$
if the rotation is purely circular, where $R_0 \equiv 8.5 $~kpc is
the distance from the Sun to the Galactic center.  The figure
shows an object which we call ``the Plume,'' centered at $\ell,b =
30\arcdeg+26\arcdeg$.  If it is at the tangent point, it has a
distance from the plane $z=3.4$~kpc. It is an irregular structure
approximately $10\arcdeg$ by $5\arcdeg$ in size with localized \HI
column densities of 2 -- 4 $\times 10^{19}$~\cmm. There are a few
nearby \HI clouds which seem to be related to the Plume, but its
main section does not seem to be a group of clouds but rather a
singular, albeit complex, object.

Figure~3 shows the entire region of our survey, and
 the Plume in the larger context of the Galactic
disk and lower halo.  Here the  \HI spectra were integrated over
60 -- 160~\kms, a range which covers the tangent point velocities
at all these longitudes.  Most of the area of Fig.~3 was observed
with the GBT, but lower angular resolution  data from the LD
survey have been added around the edges of the image. Values of
$N_{\rm HI}$ vary by several orders of magnitude from low to high
latitudes over this region, so as an aid to visualization the data
were scaled by $\sin |b|$.

The GBT data show what looks like a
system of many filamentary structures (we dub them ``whiskers'')
reaching from the disk to $b \geq 15\arcdeg$ ($z \approx 2$~kpc).
 A population of dozens of compact clouds fills the space between
the whiskers and the Plume, suggesting that the Plume itself at $ b = 25\arcdeg$
may be
 a coherent cap on top of an unusually violent eruption of gas from the
Galactic disk. Based on our inspection of the LD and WHAM survey
data (the latter is described in \S 4.2), this system
is one-sided and does
not extend below the Galactic plane.

The kinematics of the Plume is shown in relation to its spatial
structure in Figure~4. The lower part of the figure shows velocity
as a function of the Galactic longitude for the brightest \HI
peaks. Only a few clouds with velocities
significantly larger or smaller than the rest were excluded from
this Figure. We will return to them when discussing the kinematics
of the system and its surroundings. Vertical bars on each point
indicate the FWHM of the line. The upper part of the figure
relates each measurement to its position on the sky. The
measurements above or below the strip of $25\arcdeg < b <
28\arcdeg$ are filled with gray and have bolder bars in order to
distinguish them from the measurements made in the main section of
the object.

There is a clear linear dependence of the LSR
velocity of the Plume with longitude,
a relationship which extends even to the
outlying clouds. The slope of $V_{\rm LSR}(\ell)$ matches the
slope of the $^{12}$CO terminal velocities in
the Galactic plane \citep{COcurve} shown by the dashed line. We
will see that this is fully explained by the effect of projection
of Galactic rotation for an object near the tangent point and is
an indication that we are in fact dealing with a single coherent
object. The typical FWHM of the lines  is about 15 --
20~\kms, which is broader than other known halo clouds
\citep{FJL02,LP04}, suggesting that the Plume is highly
turbulent.

Examples of spectra taken at seven locations within and around the
Plume are shown in Figure~5. Although these are the brightest
lines of the system, only a few reach $T_B$~$>$~1~K. Often the
lines are not single Gaussians but have a double or even more
complex line structure. The spectrum at G28.30+30.35 is from the
highest latitude cloud we have detected -- well separated from the
main body of the Plume  but certainly part of it.

\subsection{Ionized Hydrogen} \label{ionized}

The region of our study has been observed in the \Ha\ line of ionized
hydrogen with the Wisconsin H-Alpha Mapper (WHAM; \citet{WHAM}).
Over most of the area the optical extinction is low,
and \Ha\ has been detected to a great distance from the
Sun (see also \citet{Madsen2005}).  Figure~6 shows the \Ha
emission integrated over $55 \leq V_{\rm LSR} \leq 95$~\kms,
similar to the velocity range used in Figures~2 and 3. The lower
velocity limit was chosen as optimal for avoiding the
contamination from unrelated emission, while the upper velocity is
the limit of reliable data in the WHAM survey.

The Ophiuchus superbubble is clearly a major feature in \Ha as
well as in \HI. However, unlike the \HI, which seems concentrated
in the Plume and several ``whiskers'' marking the edges of the
system, the \Ha is not limb-brightened, and if anything, is
brightest in the center of the system. We estimate that any
central cavity in the ionized gas is likely to have a radius less
than half that of the system: the \Hp is distributed over a large
volume rather than in a thin shell.

The \Ha data overlaid with \HI are shown in Figure~7. Green color
represents \Ha and purple \HI. The diagonal purple stripe at the
bottom is due to dust in the ``Great Rift'' attenuating the \Ha.
It is easy to see the correspondence of many features, e.~g., the
tips of the \HI whiskers at $(\ell, b) \approx (40\arcdeg,
15\arcdeg)$ and $(30\arcdeg, 15\arcdeg)$, clouds at $(29\arcdeg,
31\arcdeg)$ and $(22\arcdeg, 25\arcdeg)$, etc.,  but the most
spectacular is the match of the \HI Plume with the top of the
ionized hydrogen structure.  The \Ha image of some of the smaller
clouds seems shifted to a higher longitude than the \HI, but this
is likely the effect of the $1\arcdeg$ beam size of the WHAM and
the incomplete sampling of the \Ha survey. Many clouds are
represented  by only one or two pixels in \Ha (Figure~6) and much
of the emission is at the limit of the WHAM survey sensitivity.
  Additional  \Ha\ observations of this system
are now being made.

The \Ha shows the continuity between the \HI features near the
plane and the high-$z$ parts of the system. \Ha emission
associated with the largest \HI whisker continues upward and
appears to connect with the \HI Plume. This correspondence is the
primary evidence that at least some of the \HI whiskers are
related to the Plume. We conclude that we are seeing a single
system of neutral and ionized gas, with \Ha\ emission filling the
$\approx 10\arcdeg$ void between the \HI whiskers and the Plume.
This system is most probably a coherent structure of gigantic
proportions.

\subsection{Other Species}

At the moment there is no conclusive evidence that the superbubble
has been detected in anything other than \HI and \Ha.  There is
considerable structure in the soft X-ray emission as measured
by  ROSAT in this region  \citep{ROSAT}
but most of it appears to arise
 from absorption in the same dusty  foreground medium
that blocks the \Ha.  Nothing similar to the \HI or \Ha\  features
of the superbubble is found. This is not surprising if we are dealing with a
superbubble: its interior is expected to be too hot to be detected in
soft X-rays \citep{GalWinds}.

There is also no significant correlation with the radio continuum
at 408 MHz  \citep{Haslam}, which closely matches
the diffuse soft X-rays in this region. In fact, what we
see in both the radio continuum and the X-ray is the tip of the
North Polar Spur overlaid on the superbubble. This is probably
a coincidence as we can find no detailed correspondence
between the \HI and radio  or X-ray emission.  Also, the
 North Polar Spur is thought to be a local object at a
distance of a few hundred pc \citep{Bingham, Willingale} while
 we believe that the superbubble is about 7~kpc away ($\S5.1$).

Detection of UV metal resonance lines at the velocity of the superbubble
would be very useful in uncovering the origins of its gas, but to the
best of our knowledge no line of sight in a relevant direction has
ever been observed in a UV or optical absorption line.

\section{THE OPHIUCHUS SUPERBUBBLE: DERIVED PROPERTIES}

\subsection{Distance}

The \HI velocities of the superbubble, especially
the part of it near the Galactic disk, are very close to the
tangent point velocities at the corresponding longitudes (see \S5.3)
and thus we adopt the tangent point distance.  For a Sun-center
distance $R_0 = 8.5$~kpc and a nominal longitude of
$\ell=30\arcdeg$, the tangent-point distance at the base is
7.4~kpc, while if the Plume lies directly above that location, at
$b=25\arcdeg$, it must be at a distance of 8.1~kpc.

$V_{\rm LSR}$ changes slowly as a function of
distance in the vicinity of a tangent point and thus
kinematic distance estimates are not very precise
 even in the absence of non-circular
motions. We adopt a nominal distance of 7~kpc for every
part of the system and use a scaling parameter, $d_7 = d/7$~kpc to show
how derived quantities depend on the adopted distance.  For a
distance to the base of 7~kpc, the Plume, the cap on the
superbubble,  is at $z = 3.4$~kpc from the Galactic plane.

\subsection{Size and Mass}

The Plume is $10\arcdeg$ in longitude and $5\arcdeg$ in latitude,
which corresponds to $1.2 \times 0.6$~kpc at a distance of 7~kpc.
It has several concentrations
 with typical diameters  of about $100$~pc.
To estimate its mass  we integrated the column density
over the range of 55 to 100~\kms, which is a slightly larger
velocity range than is covered in Figure~2, but which includes
most of the emission. The total derived \HI mass is $3 \times
10^4 d_7^2 M_{\sun}$, and the \HI  mass of an individual clump is
about 500 -- 1000 $d_7^2 M_{\sun}$.

This mass estimate is subject to a number of uncertainties. It
does not include the lower velocity wings of a few high longitude
 clouds and it unavoidably includes some unrelated \HI from the
wings of a few low and
high-velocity sources. The former effect
is marginal but the latter one might add to the measured mass. By
choosing different velocity ranges and sky boundaries, and
measuring the mass before the stray radiation correction, we have
examined how all these factors change the estimated \HI mass.
Alternate choices of velocity range have the most influence on the
result and lead to variations in mass
of as much as a factor of 2. From a
similar analysis we derive masses of a few $10^5 d_7^2 M_{\sun}$
for each whisker, but here there is an additional uncertainty
because whiskers blend with unrelated emission at low latitudes.
The estimated mass is thus  for that part of a whisker
which lies at $b \gtrsim 4\arcdeg$. In our best estimate, the
total \HI mass of the superbubble system, evaluated as the sum of the
whiskers' masses and the mass of the Plume, is $\sim 10^6 d_7^2
M_{\sun}$.

\subsection{Kinematics}

Because of the large size of the superbubble, projection effects must
be considered in analyzing its kinematics. Thus we introduce a new
velocity coordinate, the ``deprojected'' velocity $\upsilon$.
For a point of the observed 3D space $(\ell, b, V_{\rm LSR})$,

\be \upsilon (\ell, b, V_{\rm LSR}) \equiv V_{\rm LSR} \,\sec (b)  - V_t
(\ell). \ee

\noindent Here $V_t(\ell)$ is the tangent point velocity for the
given Galactic longitude derived from the empirical polynomial
relation of \citet{COcurve},
neglecting his proposed change to the LSR,
and $V_{\rm LSR} \sec (b)$ is the
measured velocity of an \HI\ line corrected for the projection of
circular Galactic rotation with latitude.  Objects in circular
rotation near the tangent point will all have a similar value of
$|\upsilon| \sim 0$ regardless of their latitude.  This
deprojection is useful mainly in displaying and visualizing the
relationship between objects at different locations; it is less
useful for quantitative analysis.

We identified 636 \HI features which appear to be discrete clouds
in the  system and measured their central velocities.  Such
measurements have been performed everywhere it was possible above
$b = 6\arcdeg$. Closer to the Galactic plane the features are too
blended to be distinguished.  Resultant values of $\upsilon$ are
shown in the four panels of Figure~8.  The twenty points belonging
to the Plume are marked by a larger size and darker shade. They
are the same positions plotted in Figure~4.  The kinematic
coherence of the Plume stands out.  Much of the systematic variations
in $V_{\rm LSR}(\ell)$ noted in Fig.~4 arise because of the projection
effects. The variation is removed in the
deprojection (Fig.~8b).  The Plume is seen to be a distinct,
coherent structure with kinematics similar to that of the gas in
the disk below it, but  at a velocity somewhat lower than $V_t$. The
connection is explored further in $\S$~\ref{LAG}.

\subsection{Connection to Gas Nearer the Plane}

The lower part of the superbubble system shown in Figure~3 consists of
whiskers of \HI\ extending as much as 2~kpc into the Galactic
halo. The typical \HI mass of each whisker is a few $10^5\  d_7^2 \
M_{\sun}$.  The \Ha data clearly show that the superbubble has
continuity from its \HI cap, the Plume,  down to the plane,
but the existing \Ha data provide little kinematic information on
the connection.

In Fig.~8b we see some \HI features which lie spatially and
kinematically between the Plume and the bulk of the emission at
lower latitude, but their connection with the superbubble, if any,
 cannot be established from existing data.
Clouds that lie at some significant distance from the tangent
point have values of $\upsilon \ll 0$. As seen in the figure there
are many such objects within the analyzed set of measurements.
They are probably unrelated to the superbubble.

We conclude that there is a clear connection between \HI in the disk
and that in the Plume, but the kinematic structure of the entire
system remains uncertain.

\subsection{Plume Kinematics: Corotation or Lag?} \label{LAG}

An object far from the Galactic plane, like the Plume, may not be
corotating with the material below it. In fact, it is expected to
lag significantly behind Galactic rotation, i.e., $V_{\theta}(z=3\
\mathrm{kpc}) < V_{\theta}(z=0)$, and it may have  a vertical
velocity, $V_z$, as well \citep{ballist}.
We test this possibility for the Plume atop the superbubble.
As the reference for normal
Galactic rotation we use the molecular clouds
which lie close to the plane and have a small component of random
motion; their kinematics can be traced in spectral lines such as
$^{12}$CO.

Figure~9 shows  measurements of the tangent-point velocity, $V_t$,
of $^{12}$CO in the Galactic plane over the longitude range of the
superbubble \citep{COcurve} that have been fit with a spline curve which
captures the small-scale structure.  For this fit the $^{12}$CO
measurements were filtered with a $1\arcdeg$ median filter, and
evaluated every $30\arcmin$. The terminal velocity of
 $^{12}$CO changes by $>30$~\kms over this region.  The sharp
jump at $\ell \approx 35\arcdeg$ is seen in low-latitude \HI as well as CO
\citep{Burton}, and may be related to density wave streaming,
though it is not a feature of recent models \citep{Bissantz}. The
points below the CO curve show the LSR velocity of the Plume
corrected for projection of circular rotation, $V_{\rm
LSR}\,\sec(b)$, and the vertical bars show the FWHM of the \HI
lines. The CO and \HI track each other with a nearly constant
offset. The correlation between velocities in the plane and in the
superbubble cap is
shown further in Figure~10.  They  are correlated at
the 85.3~percent probability level
(Pearson correlation coefficient).

For an object near the tangent point, the velocity components
expressed in galactocentric coordinates $(R,\theta,z)$ project to
the LSR in the following way:

\be \label{VLSR} V_{\rm LSR} = R_0 \sin(\ell) \cos(b)
\left(\frac{V_\theta}{R} - \frac{V_0}{R_0}\right) + V_z \sin(b)
\ee

\noindent where $V_{\theta}$ and $V_0$ are the azimuthal
(rotational) velocities at $R$ and at the Sun, respectively, and
$V_z$ is a vertical component of motion, taken to be positive in
the direction of the north Galactic pole. Near the tangent point,
velocities which are radial with respect to the Galactic center,
$V_R$, project across the line of sight and do not enter into
$V_{\rm LSR}$.

For an object at the tangent point, where $R = R_0 \sin(\ell)$, we
can write

\be V_{\rm LSR} =  V_{\theta} - V_0\,\left( \frac{R}{R_0} \right)
\equiv V_t, \ee

\noindent and define a lag velocity

\be V_{lag}(z) = V_{\theta}(z) - V_{\theta}(0). \ee

\noindent which quantifies the difference between the rotational
velocity in the halo and that in the plane.  Rewriting
equation~\ref{VLSR} for an object at the tangent point:

\be V_{\rm LSR} = (V_t - V_{lag})\cos(b) + V_z \sin(b). \ee

We have fit this equation to the \HI in the Plume using, for $V_t$, the
spline curve from the $^{12}$CO data of  Figure~9. The solution
with $V_z \neq 0$ fails the F-test with certainty very close to
100\%, i.e., we can detect no significant vertical motion of
the Plume within the statistical error of 22~\kms.
Interpreting the difference between the Plume and CO velocities as
arising from a lag in  rotational velocity, we find $V_{lag} =
26.6\pm4.6$~\kms ($1\sigma$). The \HI data corrected for the
derived lag are shown in Figure~11. The  velocities of the \HI\
Plume atop the superbubble
 match $^{12}$CO terminal velocities in the plane extremely well, in
some cases even following fine structure in $V_t$. This is strong
evidence that the motion of the Plume is determined by the
gravitational field of the Galaxy, and not local conditions.

There is, however, another possible interpretation of the velocity
difference between the Plume and the tangent point: the superbubble
system may not be located at the tangent point. If we interpret
the Plume's velocity using a model requiring cylindrical Galactic
rotation for all heights above the disk, i.e., $V_{\theta}(z) =
V_{\theta}(0)$, its implied location  is
either at a ``near'' distance of 5~kpc (and $z=2.4$~kpc) or a
``far'' distance of 11~kpc with $z=5.4$~kpc. In this case,
however, the \HI whiskers rising up from the plane very close to
the tangent point must not be related to the Plume despite the fact
that \Ha\ emission connects all parts of the system. This alternate
interpretation still leaves the Plume extremely far from the
Galactic plane, where some lag from corotation is expected anyway.

We conclude that it is most likely that: 1) The
superbubble is a coherent system whose
kinematics derive primarily from Galactic rotation; 2) The
distance to the system is approximately the tangent point distance,
 7~kpc; 3) the kinematics of its cap, the Plume, are consistent with a lag
in Galactic rotation of $26.6\,\pm\,4.6$~\kms at a location
$R=4$~kpc and $z=3.4$~kpc; 4) The Plume shows no significant
evidence for vertical motions.

\subsection{Potential Energy}

\label{pot_energy}

As a lower limit on the energy needed to produce the superbubble,
 we can estimate the gravitational potential energy of
 the Plume.  Let us assume that
initially the  mass now in the cap was at rest in the Galactic
plane, rotating with the Galaxy. Then, using a Galactic potential
model, we can estimate the energy required for it to reach its
current position. Both the graphs in \citet{ballist} (derived from
the potential by \citet{WolfirePot}) and calculations with the
{\em GalPot} package\footnotemark ~by Walter Dehnen \citep{GalPot}
give the same result: an object in the plane at $R=4$~kpc would
have to be given a vertical velocity of $\approx 200$~\kms to
reach $z = 3.4$~kpc. Combined with our estimate for the Plume's
\HI mass this gives:

\be E_{pot} > 10^{52} \; d_7^3 \; \mathrm{erg}. \ee

An exact calculation with the {\em GalPot} package gives the
identical result of $1.8\times 10^{52}\;\mathrm{erg}$.

\footnotetext{For our purposes the differences between most of the
models by \citet{GalPot} are irrelevant. We preferred models like
their numbers 1 and 2b, which agree with the empirical rotation
curve of \citet{COcurve}.}

\subsection{Ionization}

The \Ha data can be used to derive the rate of emission of
ionizing photons illuminating the Plume from the Galactic plane.
For this purpose we use a small cloud separated from the main
body of the Plume, though clearly associated with it: the
cloud at $\ell \approx 29\arcdeg$, $b \approx 31\arcdeg$, whose
 angular size  is approximately $1\arcdeg$  in \HI.

The GBT  spectra show that this cloud has a foreground
 \NHI $= 6 \times 10^{20}$
cm$^{-2}$, implying a visual extinction $A_V = 0.3$ mag for a
standard dust-to-gas ratio \citep{DiplasSavage}.  We neglect this
modest extinction in this initial analysis of the
ionized gas, where we seek only to understand the general nature
of the system.  Subsequent studies will take this into account,
however, for it becomes increasingly important at lower latitudes.

The total \Ha intensity of this cloud in the WHAM survey is
  $I_{\alpha} \approx 0.1$ Rayleigh, which
corresponds to a production rate of $10^5$  \Ha photons cm$^{-2}$
s$^{-1}$. For Case B recombination this requires ionization by  a
Lyman continuum photon flux

\be
 F_{\rm LC} = 2 \times 10^5 \left( I_{\alpha} \over {0.1 {\rm R}} \right)
 {\rm ph\  s^{-1}}
\ee

\noindent \citep{Tufteetal98}.   Neglecting geometric factors, and
assuming that the photon source is a point located a distance $z$
from the cloud, the source production rate of Lyman-continuum
photons is

\be Q(LC) \approx 2 \times 10^6 \,I_{\alpha} \cdot 4\pi z^2. \ee

\noindent For the cloud at $29\arcdeg+31\arcdeg$ with $z = 7\
\tan(31\arcdeg) = 4.2$~kpc,

\be Q(LC) \approx 5 \times 10^{50} d_7^2 \;  \mathrm{ph} \; \mathrm{s}^{-1}. \ee

\noindent This value is consistent with the output of 100 typical
O-class stars \citep{Ostars}.

The Ophiucus superbubble lies near many H\,II regions and young stellar
clusters, though its size is so large that we cannot pinpoint a singular
source of its ionization.  The W43 cluster at $\ell \approx
30\arcdeg$ generates $10^{51}$ Lyman-continuum photons s$^{-1}$
\citep{Smith}, enough to ionize the Plume if the path between the
two is unobscured.  This is discussed further in $\S7$.

\subsection{\Hp Mass of the Superbubble}

Assuming that the temperature of the
\Hp is 8000~K, a typical value for the Galactic ionized medium
\citep{ReynoldsT8000}, each Rayleigh of \Ha\ emission corresponds
to an emission measure of 2.25~cm$^{-6}$~pc \citep{WIM}. The superbubble
is about 2~kpc wide and we assume the same value
 for the emission depth $l_{EM}$ (which is also subject to
the distance uncertainty factor $d_7$). The average electron
density inside the structure is

\be \bar{n}_e\,(\mathrm{cm}^{-3}) = \left( \frac{2.25
 I_{\alpha}
(\mathrm{R})}{f\, l_{EM} (\mathrm{pc})\, d_7 } \right)^{1/2} \,
\ee

\noindent where $f$ is the filling factor of ionized gas along the
line of sight. The lack of strong limb-brightening in \Ha
($\S$\ref{ionized}) leads us to adopt $f \ga 0.5$. For simplicity
we ignore a possible difference between the line-of-sight and
volume filling factors. In the superbubble the typical
observed $I_{\alpha} = 0.2$~R, so, neglecting extinction,
$\bar{n}_e = 0.015\, f^{-1/2}\, d_7^{-1/2}$~\cmmm. The \Hp mass is
given by

\be M_{\mathrm{H}^+} (M_{\sun}) = 2.47\times 10^7 f\, \bar{n}_e
(\mathrm{cm}^{-3}) \,V(\mathrm{kpc}^{3}). \ee

\noindent For the 5 -- 8~$d_7^3$~kpc$^3$ volume of the system, the
values of $\bar{n}_e$,  and $f$   give \be M_{\mathrm{H}^+} = 1 -
3 \times\,10^6\,d_7^{2.5}\,M_{\sun}, \ee

\noindent a mass similar to that in \HI.

\subsection{Vertical Density Structure of a Whisker}

The vertical  structure of a ``whisker'' can give insight into its
origin. Figure~12 shows the best example of an \HI whisker which
is likely to be connected to the superbubble. Here the emission is
integrated over $70 \leq V_{\rm LSR} \leq 90$~\kms, which covers
the tangent point velocities at its longitude.  There are a number
of similar \HI\ features in the GBT  data.    Figure~13 shows the
whisker's $N_{\rm HI}$ averaged over $5\arcdeg$ in longitude. For
comparison, the solid curve is the \NHI$(z)$ expected for  a
1.6~kpc path through the Dickey-Lockman (DL) empirical \HI layer
\citep{FJL84,DLHI}.  The DL function was derived from 21~cm
measurements covering  both sides of the Galactic plane and
averaged over  $3.5 \le R \le 7.5$~kpc.  It should be
representative of the vertical structure of the \HI layer in the
inner Galaxy.

We see that the vertical density structure of the whisker
resembles that of the average interstellar medium.  This would not
necessarily be the case if the whisker were a column of gas thrust
up from the disk.  Indeed, a similar analysis along a cut through
the Plume shows it as a clear excess of gas above any scaled DL
curve at $z=3.4$~kpc. Thus we have the strong implication that the
whisker is gas swept up from the side, e.g., perpendicular to the
walls of an expanding bubble. The gas in the Plume has been
carried up to its location, but the gas in the whisker has not. If
we could separate the whisker from unrelated emission, its
effective depth, i.e., the path through the DL layer needed to
give its \NHI, would be a measure of the volume swept out to make
the whisker.  The value of 1.6~kpc for the curve in Figure~13 is
consistent with the size of the system, but should not be
given much significance at this stage of our understanding. This
particular whisker is seen in \Ha, but is so faint that
 a more detailed analysis of its ionization is impossible.

\HI ``worms'', objects morphologically similar to our whiskers,
have been discussed by \citet{Heiles84} who suggested that they
are ``probably parts of shells that are open at the top''. This is
consistent with our conjecture that whiskers are formed by a
sideways motion rather than an upward thrust.

\section{A MODEL FOR THE OPHIUCHUS SUPERBUBBLE}

It is possible to test the hypothesis that this entire system is a
superbubble blown by a cluster of young stars in the disk by
comparing our data with a  model of a superbubble. We use the
idealized analytical Kompaneets 2D model \citep{Kompan60} to draw
possible bubble boundaries, determine the position of the walls
and cap in relation to the parent star cluster, and derive a
general estimate of the time and energy scales required in such a
scenario.  The Kompaneets model has been adopted by
\citet{MacLow}, \citet{Bisnovatyj} and then by
\citet{KompanBubble} to describe a superbubble expanding from the
disk of the Galaxy into an exponential atmosphere with a scale
height $H$. The main virtues of such a simple model are its ease
of use and clear representation of the principal physical
processes involved. It also has many limitations. It is not as
exact as modern 2D and 3D simulations and, as noted by
\citet{MacLow89} and \citet{KooMcKee0}, its numerical predictions
may be off by up to a factor of 2. It also describes the bubble
evolving in complete isolation while in reality its evolution is
influenced by many external factors like Galactic gravitational
and magnetic fields, perturbation by other bubbles, etc.
Nevertheless, the model is a suitable starting point for checking
the plausibility of the superbubble hypothesis and its
quantitative errors are acceptable given
 that all the properties of the Ophiuchus system are as yet known
only to within a factor of a few.

The model follows the standard paradigm of
adiabatically evolving bubbles \citep{Weaver, MacLow}. It assumes
that the bubble is formed around a cluster of stars with a
constant total mechanical luminosity $L_0$. Initially the cluster
winds expand freely and form a shock driving into the ambient ISM.
Almost from the time of the bubble's formation it can be treated
as a very thin and dense shock-shell filled with a rarefied gas.
Assuming that the shock front moves normal to itself and that its
expansion speed is determined only by the internal pressure
$P(t)$, ambient density $\rho(z)$, and the ratio of the specific
heats $\gamma$, \citet{Kompan60} found that the shell's shape at
every moment of time is described by a curved surface from the
following family:

\be \label{KompanCurve} r(z,y) = 2H \arccos \left[ \frac{1}{2}
e^{z/2H} \left( 1 - \frac{y^2}{4H^2} + e^{-z/H} \right) \right].
\ee

\noindent Here $r$ is the radial cylindrical variable: as the
density is assumed constant for fixed $z$ the surfaces are axially
symmetric. The top and bottom of the curve, where $r = 0$, are
located at: \be z_{1,2} = -2H\,\ln\left( 1 \mp \frac{y}{2H}
\right).\ee

The parameter $y$ has dimensions of length and varies from 0 to
$2H$. It is a nonlinear function of time, energy and volume of the
bubble, but its meaning is best understood from a purely
geometrical point of view. It follows from
equation~\ref{KompanCurve} that by using $r/2H$, $z/2H$ and $y/2H$
instead of $r$, $z$ and $y$ the equation can be rendered
dimensionless.   The scale height $H$ then sets the scale of the
shell, while its shape is governed solely by the parameter $y$.
The  dimensionless ``evolution factor'' $y/2H$ varies from 0 to 1
describing the evolutionary stage of the bubble. A value of zero
corresponds to a point at the source at the moment of origin,
$t=0$. For $y/2H \lesssim 0.5$ the shape is almost spherical, then
for $0.5 \lesssim y/2H < 1$ it becomes  ellipsoid-like, getting
more and more elongated in the $z$ direction. Finally, for
$y/2H=1$ the bubble's surface is a paraboloid-like shape
stretching up to infinity. Physically this means infinite shock
acceleration in the upward direction due to a strong density
gradient, the ``blowout'' scenario, when the bubble's top is
completely disrupted and the bubble becomes a ``chimney''.

Another possibility is the ``stall'' scenario. For values of $y
\approx 2H$ the strong shock approximation is no longer valid and
thus the Kompaneets model becomes unphysical. The shell expansion
stalls when its speed falls to the sound speed of the external
medium, which is the same moment when the pressure in and out of
the bubble equalizes. By then the walls and the cap of the bubble
begin to decompose and finally the shell merges with the
surrounding ISM.

Ideally, to allow for the development of a one-sided bubble its
origin must be somewhat above the Galactic plane. In reality, even
if this condition is not satisfied the development of the downward
lobe can be blocked by a dense molecular cloud or some other
density fluctuation common in the Galactic plane, especially in a
spiral arm where the bubble's parent star cluster was likely
located.

Within the Kompaneets framework,  the age and internal thermal
energy of the bubble, $E_{th}$, (which is just the mechanical
energy from the source cluster, $L_0 t$, minus the work done on
the shell expansion) is obtained by solving the equations
numerically to get $y$ explicitly through time. However,
\citet{KompanBubble}, who performed this numerical evaluation,
found that the solutions for time and energy closely follow those
for the much simpler spherically symmetric model of
\citet{SpherBubble} if their shell radius $R_s$ is made equal to
$y$. Thus we can use the following approximate formulae for the
age of the bubble and its internal energy
\citep{SpherBubble,KompanBubble}:

\begin{eqnarray} t & \simeq & \left(\frac{154 \pi}{125}\right)^{1/3} \: y^{5/3} \:
\rho_0^{1/3} \: L_0^{-1/3}, \label{age}\\
E_{th} & \simeq & \frac{5}{11} \: L_0 \, t. \label{th_energy}
\end{eqnarray}

\noindent Note that when the shell has evolved to its maximum
diameter the Kompaneets energy is less than the one calculated
from the spherical model.

In Figure~14 two Kompaneets curves are plotted on the combined \HI
and \Ha data from Figure~7. We have assumed a constant distance
scale (shown in the upper left corner) for the entire figure based
on the distance to the system of 7~kpc.  Both models have
their center of origin at the same $z$, close to the Galactic
plane. The model parameters are given in Table~1.

The solid curve (Model~1) was fit by hand to the brightest \HI
whiskers and clouds, and goes through the dense \Ha region at
$\ell \approx 35 \arcdeg$ , $15\arcdeg \la b \la 25\arcdeg$. In
order to do this the curve had to be tilted by $14.5\arcdeg$ in
the outward direction relative to the Galactic center. This tilt
is consistent with the behavior of the whiskers, which are all
tilted by $10\arcdeg - 20\arcdeg$ in the same direction. It is
also consistent with the effect of the Galactic gravitational
potential which the Kompaneets model does not take into account at
all: any object thrown vertically from the Galactic plane at $\ell
\sim 30\arcdeg$ will drift to a higher longitude while it moves
upward. The results of throwing an object to reach $z$ of a few
kpc would give us the same effective outward tilt angles of
$10\arcdeg - 20\arcdeg$ (\citet{ballist}, and calculations from
{\em GalPot} \citep{GalPot}). Model~1 also has a benefit of being
narrow,  and because of that does not extend significantly into
and below the Galactic plane.

A wider Kompaneets model that still reaches the same height from
the plane would need either an origin at a higher $z$, or an
extent some distance below the Galactic plane where, in a
realistic situation, it would create a second lobe extending into
the halo below the disk.  We do not observe a second lobe, but
created Model~2 (the dashed curve) mainly to test the generality
of the results derived from Model~1.  Model~2 is not tilted and
was manually fit to contain most of \HI and \Ha of the system.

Both models have an evolution factor, $y/2H$, almost equal to
unity.  It is interesting that the atmospheric scale height needed
to match the shape and size of the system is not unreasonable. The
position of the cap in relation to the Galactic disk where the
bubble's source is located is consistent with the superbubble
hypothesis. These conclusions would hold for any Kompaneets curve
fit to the Ophiuchus system  because the height of the superbubble
is so much larger than its width. The meaning is simple: this
bubble is already stalled/blown and is dissipating, or it is
approaching the stall/blowout. This is the main qualitative result
we get from the Kompaneets model.

We evaluated eqs.~\ref{age} and
\ref{th_energy} for arbitrary $n_0$ and $L_0$ and then for a few
typical values. A typical young star cluster has a
mechanical luminosity
of $10^5 \: L_{\sun}$ \citep{DoveShullFerrara, nsmith06}
so this is used as a  reference value.
 The central density is  chosen to be 1 \cmmm, though other
possibilities were examined to understand the sensitivity
of the derived properties to the
input density.  The results are presented in  Table~1. One column
shows  how the calculations would change depending on the value of
the distance to the system. As the bubble is close to the end of
its evolution, the age calculation should be treated as a lower
limit, while the energy is an upper limit.

The model properties differ by a factor of only 2.5, an
insignificant difference in view of its limitations.
Any Kompaneets model constrained to have a source near the plane
and upper end at the Plume would give similar results.  The models
give an age of the bubble whose order of magnitude is: \be t \ga
10 \: d_7^{5/3} \; \mathrm{Myr},\ee \noindent which can be
compared with the ballistic age of the Plume, i.~e., the time
necessary for it to reach its current position at
$z\approx3\;\mathrm{kpc}$ from a single vertical thrust of
velocity applied at the Galactic plane.  From the {\em GalPot}
package this time was found to be $\approx 30\;\mathrm{Myrs}$. It
is interesting that this age corresponds to the time of onset of
fragmentation of a superbubble shell due to instabilities
\citep{DoveShullFerrara}. Both Rayleigh-Taylor and gravitational
instabilities may appear in the late phases of bubble evolution.
When a bubble has expanded to the point that its interior pressure
is similar to the exterior pressure, characteristical
Rayleigh-Taylor ``fingers'' will form and break, producing a
debris of cold cloudlets \citep{Breitschwerdt06}. The shell may
also become gravitationally unstable at the same time as well, also
resulting in shell fragmentation
\citep{ElmegreenLada77,McCrayKafatos87,Voit88}. The structure of
the Plume suggests that it might be just at this point in its
evolution. Based on coincidence of these three different age
estimates we conclude that the age of the bubble is most likely to
be $\approx 30$~Myr.

The total  internal energy of the system has the limit

\be E_{th} \la 10^{53} \: d_7^{5/3} \; \mathrm{erg}, \ee

\noindent which is comparable to the estimate of the gravitational
potential energy of the Plume discussed in \S~\ref{pot_energy}.
Combining these two estimates, the total energetics of the
system is in the $10^{52 - 53}\;\mathrm{erg}$ range. Of course, we
do not know the nature of the parent stellar cluster,
 so we cannot tell if it suffers the ``energy-deficit
problem'' observed in other systems \citep{OeyGarcia,Cooper04}.
The Ophiuchus superbubble  could have been formed by an OB
association containing about 70 O stars, similar to the Carina
Nebula, which will contribute to the ISM about
$2.6\,\times\,10^{52}$~erg through stellar winds over 3~Myr
\citep{nsmith06}.

\section{THE OPHIUCHUS SUPERBUBBLE IN THE GALAXY}

The \HI and \Ha superbubble lies above a section of the Galaxy
which contains many sites of active star formation.  The H\,II region
and molecular cloud complex W43, at $\ell \approx 30\arcdeg$, near
the tangent point of the Scutum spiral arm, has been called a
`mini-starburst', where the  star formation efficiency
has apparently been enhanced over the last $10^6$ yr
\citep{Motte}. The brightest H\,II region in W43 is G29.944-0.042
with a radio recombination line velocity of 96.7~\kms, and there
are many other H\,II regions within a few degrees that have the
$70 - 100$~\kms  velocities of the \HI whiskers and Plume
\citep{FJL89}.  Although the superbubble has an age $>10$~Myr and
was likely produced by a generation of young stars previous to the
current W43 cluster, W43 is known to generate at least $10^{51}$
Lyman-continuum photons s$^{-1}$ based on the radio observations
that trace the absorbed fraction \citep{Smith}. The W43 stars
clearly have sufficient ionizing luminosity to ionize the superbubble,
if the disk medium allows a moderate leakage of Lyman-continuum
photons into the halo.

The Plume itself is so far from the plane that it is probably
exposed to ionizing radiation from several spiral arms in the
inner Galaxy. Because we have a good estimate of its distance, it
can act as a probe of the radiation field above the disk.  It has
an \Ha flux similar to that of some high-velocity clouds also
detected with the WHAM instrument
 \citep{Tufteetal98,Tufteetal02}, and the implied
Lyman-continuum photon flux needed to maintain its ionization, $2
\times 10^{5}$~photons~cm$^{-2}$~s$^{-1}$, is in the range of that
expected from models of an object at its location
\citep{BlandHawthornMaloney,Putman}.

\section{CONCLUDING DISCUSSION}

Using 21~cm \HI spectra measured with the GBT, we have discovered
a large coherent structure located in the inner Galaxy at a
distance of about 4~kpc from the Galactic center and 7~kpc from
the Sun. Its top reaches $>4$~kpc above the Galactic plane and is
visible in both ionized and neutral hydrogen emission.  The
structure of the \HI, the location and
intensity of the \Ha emission, and the analysis of the system's
kinematics, give a consistent picture: most probably we are seeing
a superbubble blown by the joint action of stellar winds and
multiple supernovae from a star cluster in one of the Galaxy's
spiral arms. The model shows that the energetics required to power
the creation of a bubble of this size is of the same order as is
produced from a typical OB association. The structure's age is
$\approx 30$~Myr and the OB stars in the cluster, which formed the
bubble, have thus already evolved off the main sequence. A
different, younger cluster must be the
source of ionizing photons which produce the observed \Ha.
 A summary of the properties of
the \HI  cap on the system is given in Table~2, and a
summary of the properties of the system as a whole is in Table~3.

The Plume, the neutral cap  on top of the system,  is  very
irregular  with broad lines suggesting substantial turbulent
motions.  And yet, its overall kinematics match  the kinematics of
molecular clouds in the plane below it quite well, with a lag of
27~\kms. Extra-planar  gas is expected to show a gradient in
rotational velocity arising from a change in the gravity vector
with $z$, and recent models have attempted to  reproduce the
magnitude of the effect observed in other galaxies
\citep{ballist,Barnabe,Fraternali}.  In the Milky Way, evidence
for deviations from corotation is suggestive but not compelling
\citep{Savage90,SavageSemLu}.  The Plume stands as the best single
example of a Galactic cloud with a significant, and coherent, lag
behind corotation.

The Kompaneets model indicates that we are seeing either the late
stages of the bubble's development or the early stages of its
decomposition. This is consistent with our failure to detect any
significant vertical motion of the Plume.  At the age of this
system several instabilities should begin to fragment a
superbubble shell \citep{DoveShullFerrara}. The turbulent,
irregular structure of the Plume, with its outlying clouds, may be
a sign that this process is already underway.  In about 30~Myr
(the ballistic free-fall time) all the material will return
to the Galactic disk.

The vertical density structure of one of the \HI whiskers
 found at the base of the system is similar to the
average vertical density structure over the inner Galaxy.  This is
consistent with the hypothesis that this  whisker is part of the
superbubble's walls swept by an expansion perpendicular to its
surface. Despite its suggestive appearance, we do not believe that
this whisker results from an outflow along its axis but rather a
sidewise push.  Other vertical structures have been identified in
Galactic \HI and interpreted as outflows
\citep{Heiles84,Koo,English,Avillez,Asgekar,KudohBasu2004}.
Although the mass of some of these is in range of the mass of the
whiskers in the Ophiuchus superbubble system, they are typically
smaller by a factor of ten in size. The whiskers detected here
more nearly resemble the vertical dust lanes seen with the WIYN
telescope in NGC 891  as pillars of extinction extending to $z
\sim 2$~kpc against the light of that galaxy \citep{Howk}. Like
the whisker of Fig.~12, they contain $\sim 10^5 M_{\sun}$.  It is
interesting that the linear resolution of the GBT in the 21~cm
line of \HI at the Ophiuchus superbubble is essentially identical
to that of the WIYN telescope at NGC 891.

This superbubble is very different from an M82-type nuclear
starburst \citep{Weiss,Matsushita}: the energies and densities
involved are orders of magnitude smaller.  It can, however,
be compared with known bubbles in
normal and dwarf galaxies.

An analysis of observations of bubbles and shells both in the
Milky Way and other galaxies shows that bubbles fall into two
general categories divided by their age. ``Young'' bubbles have
typical sizes of a few hundred pc, expansion velocities of $\ga
20$~\kms and ages of a few Myr.   ``Old'' bubbles have typical
sizes of more than 1~kpc, expansion velocities of $\la 10$~\kms
and ages of a few tens of Myr.  Old supershells usually are not
very abundant in normal spiral galaxies, possibly because of the
presence of differential rotation, destructive to large coherent
structures. However there are still several known in the Milky Way
and a few in similar galaxies like M31 and M33
(\citet{McClure-Griffiths,Kim,Ehlerova,WalterBrinks} and
references therein).  In the shell classification scheme of
\citet{Kim}, which is based on the relation between \HI and \Ha,
the Ophiuchus superbubble belongs either to Type I (shell filled
with ionized gas) or Type V (discrete H\,II regions inside the
shell due to secondary star formation inside the shell). Type I is
a characteristic of young bubbles at the earliest stages of their
development so it probably does not apply here, while Type V is
consistent with both the old age and a large number of H\,II
regions near the base of the superbubble, and also with the broad
distribution of \Hp within the system.  New ionization sources
develop after the death of the parent cluster not by chance, but
as a consequence of the bubble's evolution.

The Ophiuchus superbubble with its size  $\sim 2$~kpc, age of
$\approx 30$~Myr and lack of detectable expansion seems to be a
typical old superbubble. Its total mass of a few $10^6 \,
M_{\sun}$ and energy of $10^{53}$~ergs are also typical of bubbles
of this size. But its large size is unusual for the small
galactocentric distance of just 4~kpc: it is at least twice as
large as any known \HI  bubble at a similar location in the Galaxy
(see Fig.~16 in \citet{McClure-Griffiths}). With its radius larger
than the \HI scale height this bubble possibly belongs in a
special class of events \citep{OeyClarke}. Still, creation of such
superbubbles should be commonplace in Galactic spiral arms, so
other old superbubbles similar to this one probably exist in the
Milky Way. It may not be easy, however, to detect them. There is
no expansion or vertical motion to provide an easily recognizable
velocity signature.  If a similar object were far from a tangent
point, it would blend with local gas and  be almost undetectable
due to its low column density. Finally, one needs a nearby
independent younger star cluster to illuminate it after the parent
cluster is dead in order to produce \Ha.

Our understanding of this unique system is just beginning. We
should emphasize that the \HI column density of the structures
discussed in this paper often is so low that their detection was
only possible through the unique combination of the high
sensitivity, spatial, and spectral resolution of the GBT. Even so,
we may be detecting only the brighter HI peaks in this system and
missing faint, diffuse emission. The Plume, for example, which
appears in Fig~2 as a number of distinct parts, may be a single
structure connected with a thin envelope. In the absence of more
sensitive \HI observations many interesting aspects of this system
must remain unknown. The analysis of its ionized component, in
particular, is crude, as the angular resolution of the \Ha
measurements is poor and our assumptions about the geometry of the
system are primitive.  A more sophisticated analysis, using higher
angular resolution \Ha data and including a detailed estimate of
foreground extinction, would be most rewarding. Measurements in UV
absorption lines through this system would allow us to study its
internal structure as a function of ionization and search for
abundance anomalies indicative of the enrichment which accompanies
supernova-driven bubbles.

\acknowledgments

\section{Acknowledgments}

We thank Matt Haffner, Greg Madsen and Ron Reynolds of the WHAM
group for their help and advice about ionized gas.  Walter Dehnen
assisted us with his {\em GalPot} package. Y. P. thanks his wife
Elena Plechakova for her help in the arduous task of processing
the kinematic measurements of the Plume. We also thank
Carl Heiles for his attention to this paper and an interesting
discussion of its results, and an anonymous referee whose
comments allowed us to improve this paper in a number of ways.  The research of Y.
P. at NRAO was supported by NRAO predoctoral fellowship.  The
Wisconsin H-Alpha Mapper is funded by a grant from the National
Science Foundation.

\clearpage

\begin{deluxetable}{l c c c}
\tablewidth{0pt} \tablecaption{Kompaneets Model}

\tablehead{ \colhead{Property} & \colhead{Distance
factor\tablenotemark{a}} & \colhead{Model 1} & \colhead{Model 2} }

\startdata

{\bf Model parameters:} &  &  & \\
Evolution factor, $y/2H$ & - & 0.999 & 0.980 \\
Scale height $H$, kpc & $d_7$ & 0.23 & 0.42 \\
{\bf Results:} & & & \\
Age $t$, Myr, as a function & & & \\
of source luminosity $L_0$ (in $L_{\sun}$) & & & \\
and source density $n_0$ (in \cmmm)& $d_7^{5/3}$ & $680 \;
n_0^{1/3} L_0^{-1/3}$& $1760 \;  n_0^{1/3} L_0^{-1/3}$\\
&&&\\
$L_0=10^5 L_{\sun}$, $n_0 = 1$ \cmmm &  & 15 & 38\\
&&&\\
Internal energy $E_{th}$, erg & & & \\
 as a function
of $L_0$ and $n_0$ & $d_7^{5/3}$ & $4 \times 10^{49} \;
 n_0^{1/3} L_0^{2/3}$& $10^{50}  \; n_0^{1/3}  L_0^{2/3}$\\
 &&&\\
$L_0=10^5 L_{\sun}$, $n_0 = 1$ \cmmm &  & $8 \times 10^{52}$ & $2
\times
10^{53}$\\

\enddata

\tablenotetext{a}{The factor $d_7$ is introduced in \S 5.1. For a
distance different from 7~kpc each table value should be
multiplied by the corresponding power of $d_7$.}

\end{deluxetable}

\begin{deluxetable}{l c c c}
\tablewidth{0pt} \tablecaption{The Properties of the Plume}

\tablehead{ \colhead{Property} & \colhead{Distance factor} &
\colhead{Unit} & \colhead {Value} }

 \startdata

Distance & $d_7$ & kpc & 7 \\
Height above the Galactic plane & $d_7$ & kpc & 3.4 \\
Size & $d_7$ & kpc & 1.2 $\times$ 0.6 \\
\HI Mass & $d_7^2$ & $M_{\sun}$ & $3 \times 10^4$ \\
Characteristic LSR velocity & - & \kms & 70 \\
Typical FWHM & - & \kms & 15 \\
Potential energy & $d_7^3$ & erg & $10^{52}$ \\
Ionization rate & $d_7^2$ & photons s$^{-1}$ & $5 \times 10^{50}$ \\

\enddata

\end{deluxetable}

\begin{deluxetable}{l c c c}
\tablewidth{0pt} \tablecaption{The Properties of the Entire Plume
System}

\tablehead{ \colhead{Property} & \colhead{Distance factor} &
\colhead{Unit} & \colhead {Value} }

\startdata

Distance & $d_7$ & kpc & 7 \\
Size & $d_7$ & kpc & 2.7 $\times$ 4.2 \\
\HI Mass & $d_7^2$ & $M_{\sun}$ & $\sim 10^6$ \\
\Hp Mass & $d_7^{2.5}$ & $M_{\sun}$ & $1 - 3 \times\,10^6$ \\
Age & $d_7^{5/3}$ & Myr &  $\approx 30$\\
Thermal energy & $d_7^{5/3}$ & erg & $10^{53}$ \\

\enddata

\end{deluxetable}

\clearpage

\begin{figure*}
\plotone {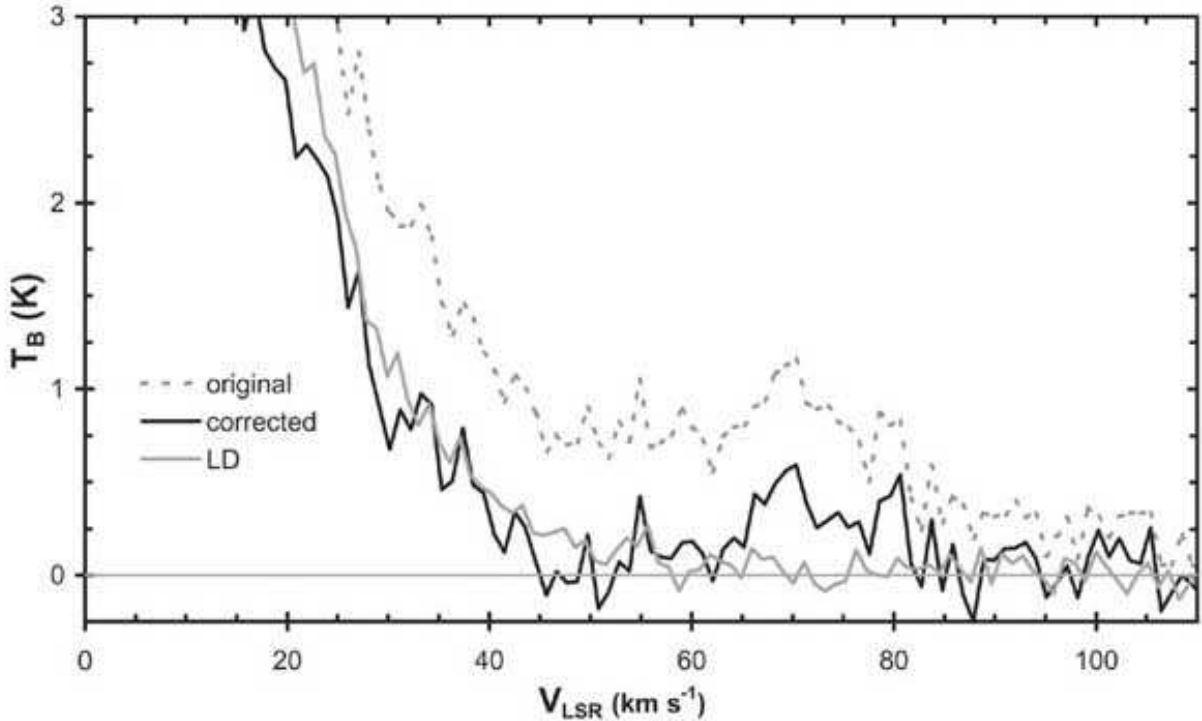} \caption{Stray radiation correction. The dashed
gray line shows an example of a GBT \HI spectrum (at $9\arcmin$
resolution) observed
 at $\ell, b = 28\fdg 70, 20\fdg 35$.
The solid gray line is the \HI spectrum from the closest point,
 of the Leiden-Dwingeloo (LD) $36\arcmin$ resolution \HI
survey at  $\ell, b = 28\fdg 50, 20\fdg 50$
 \citep{LD}. The solid black line is
the final spectrum, corrected by subtracting from the original GBT
spectrum the difference between the GBT data (convolved to
match the LD beam) and LD spectrum.  The final GBT spectrum shows emission
at $V_{\rm LSR} \approx 70$ km s$^{-1}$ from a compact cloud which
is not detectable in the LD data.}
\end{figure*}

\clearpage
\begin{figure*}
\plotone{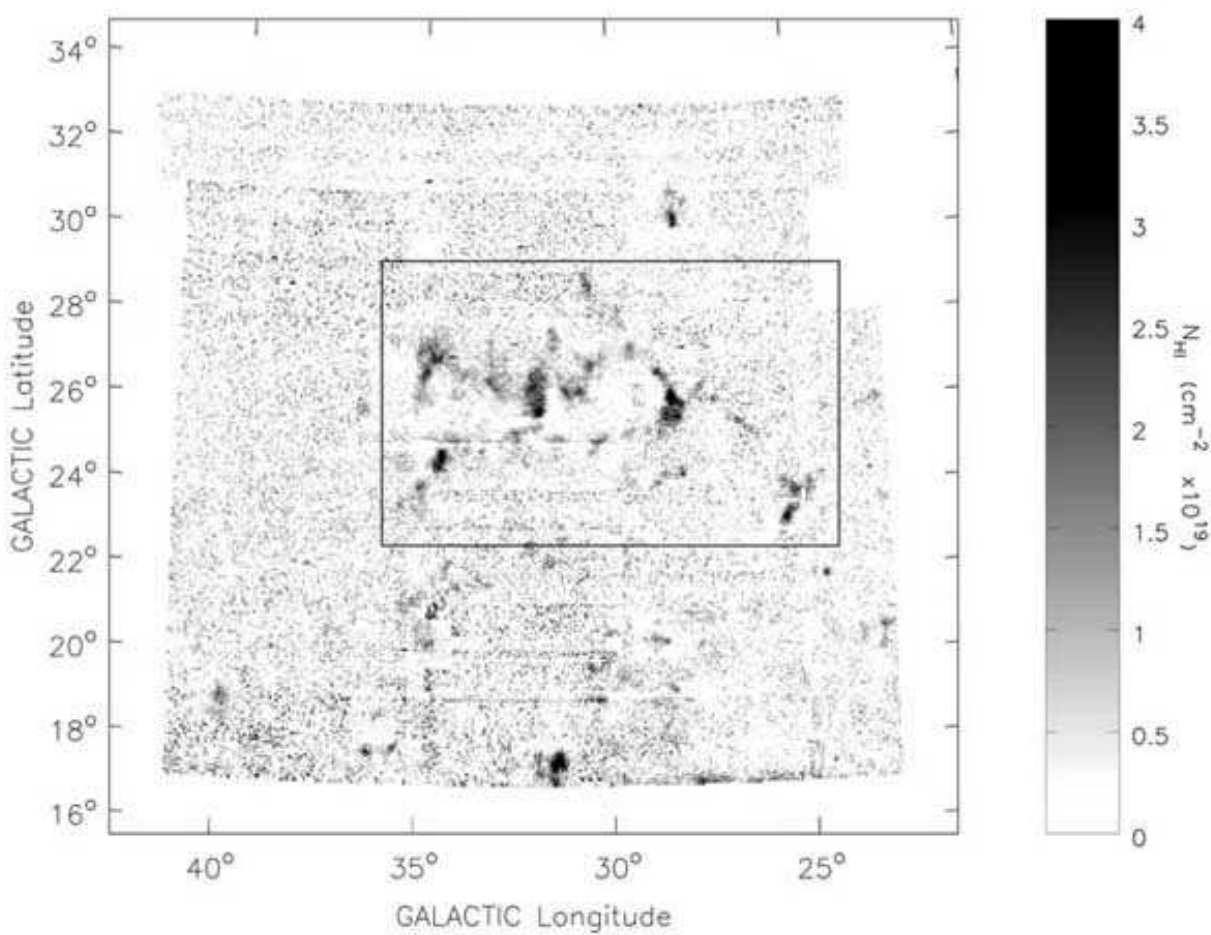} \caption{``The Plume'' in \HI.
 An \HI column density map derived from GBT spectra integrated over
$60 \leq V_{\rm LSR} \leq 100$~\kms. The figure shows an object
which we call ``the Plume.''  It has peak \HI column densities of
2 -- 4
 $\times 10^{19}$~\cmm. The main part of it lies within the box, but
 there are a few separate clouds nearby with similar kinematics which
 are likely part of it.  The cloud at $(\ell,b) \approx (29\arcdeg,
 31\arcdeg)$ is the highest-latitude member of this group discovered
 so far.}
\end{figure*}

\clearpage
\begin{figure*}
\plotone{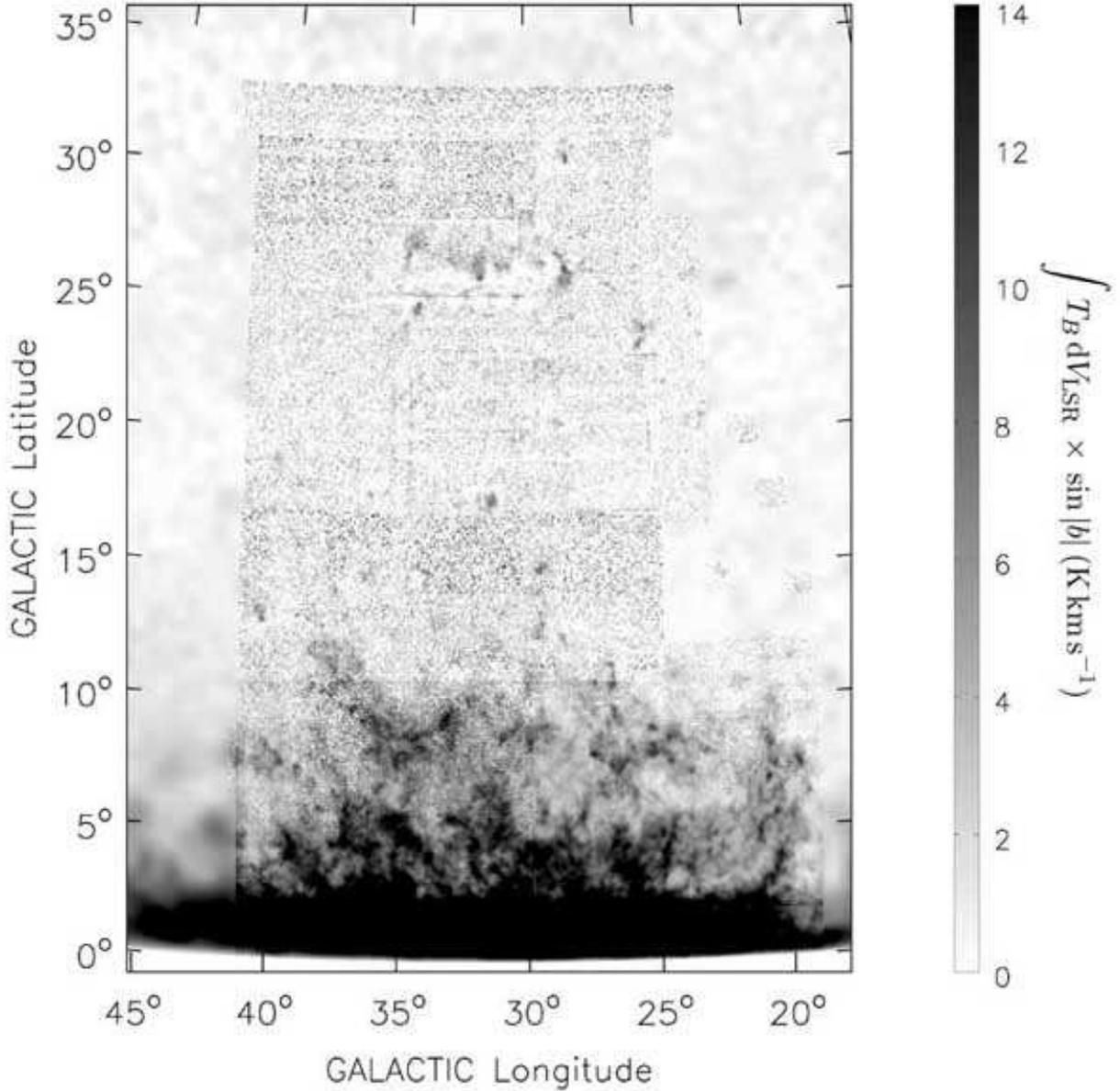} \caption{The Ophiuchus superbubble in \HI.
Spectra integrated over $60 \leq V_{\rm LSR} \leq 160$~\kms\ show
the Plume in the larger context of the Galactic disk and lower
halo. The integration covers all the tangent point velocities at
these longitudes. GBT \HI spectra are used for most of the map,
but data from the
 lower resolution Leiden-Dwingeloo survey are used around the edges.
The intensity scale has been multiplied by $\sin |b|$ to compress
the dynamic range and show structures more clearly. Several
coherent \HI features (we dub them ``whiskers'') stretch from the
disk to  $b \geq 15\arcdeg$ ($z \approx 2$~kpc). Dozens of compact
but relatively dense clouds fill the space between the whiskers
and the Plume. The Plume itself appears to  be a cap on top of an
unusually violent eruption of gas from the Galactic disk.}
\end{figure*}

\clearpage
\begin{figure*}
\plotone{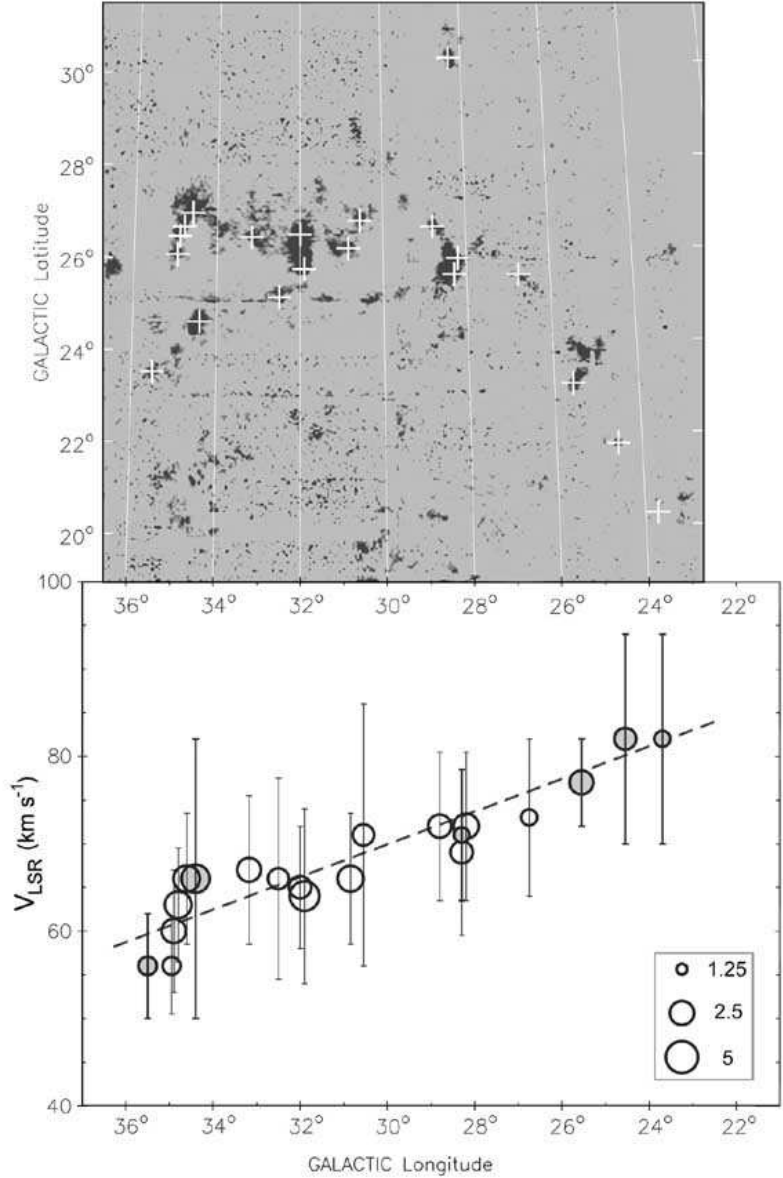} \caption{Kinematics of the Plume in relation to
its spatial structure. The lower part of the figure shows the
velocity of all of the clearly discernable features of the Plume
(marked by crosses in the upper panel).
 The area of each circle is proportional to $\log(N_{\mathrm{H\,I}})\, - 19$, where
\NHI is the column density in \cmm;  the legend shows circle sizes
for a few values of \NHI in units of $10^{19}$~\cmm. The bar
through each point shows the FWHM of the line.
Measurements made above or below the main section of the
Plume at $25\arcdeg < b < 28\arcdeg$ are filled with gray and have
bolder bars, and show that the coherence of the structure extends
even to outlying clouds.  The dashed line is
a linear fit to the CO terminal velocity measurements in the
Galactic plane \citep{COcurve} with an offset discussed in
$\S5.5$.  It shows that the linear dependence of LSR velocity with
longitude is fully explained by Galactic rotation and projection
effects.  }
\end{figure*}

\clearpage
\begin{figure}
\plotone{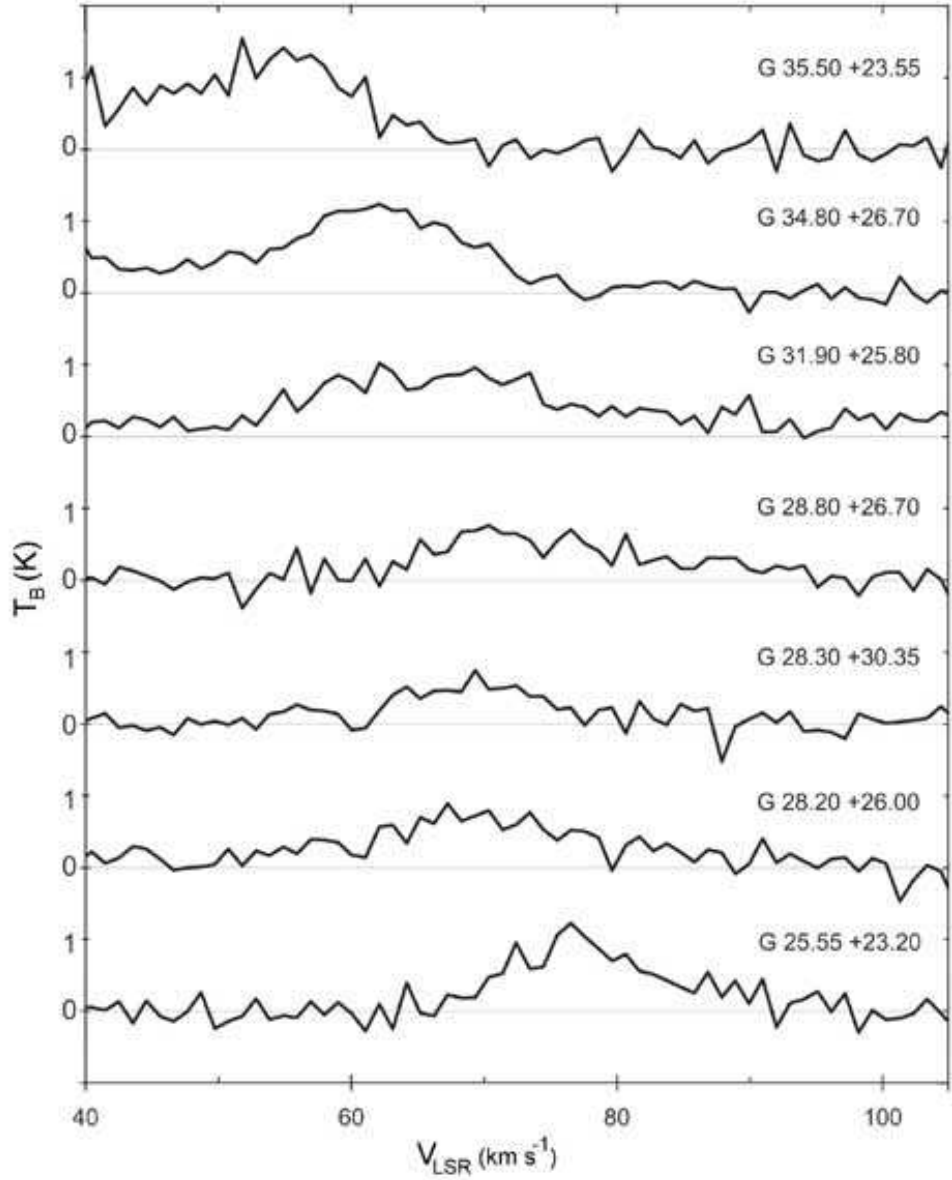} \caption{Examples of GBT \HI spectra taken at
seven locations within and near the Plume. The typical emission
FWHM is 15 -- 20~\kms, which is broader than usual for halo
clouds, suggesting high turbulence. Unusually wide wings imply a
double Gaussian or even more complex line structure. The cloud at
G28.30+30.35 lies considerably above the main body of the Plume,
but its spectrum is similar to the others. }
\end{figure}

\clearpage
\begin{figure*}
\plotone {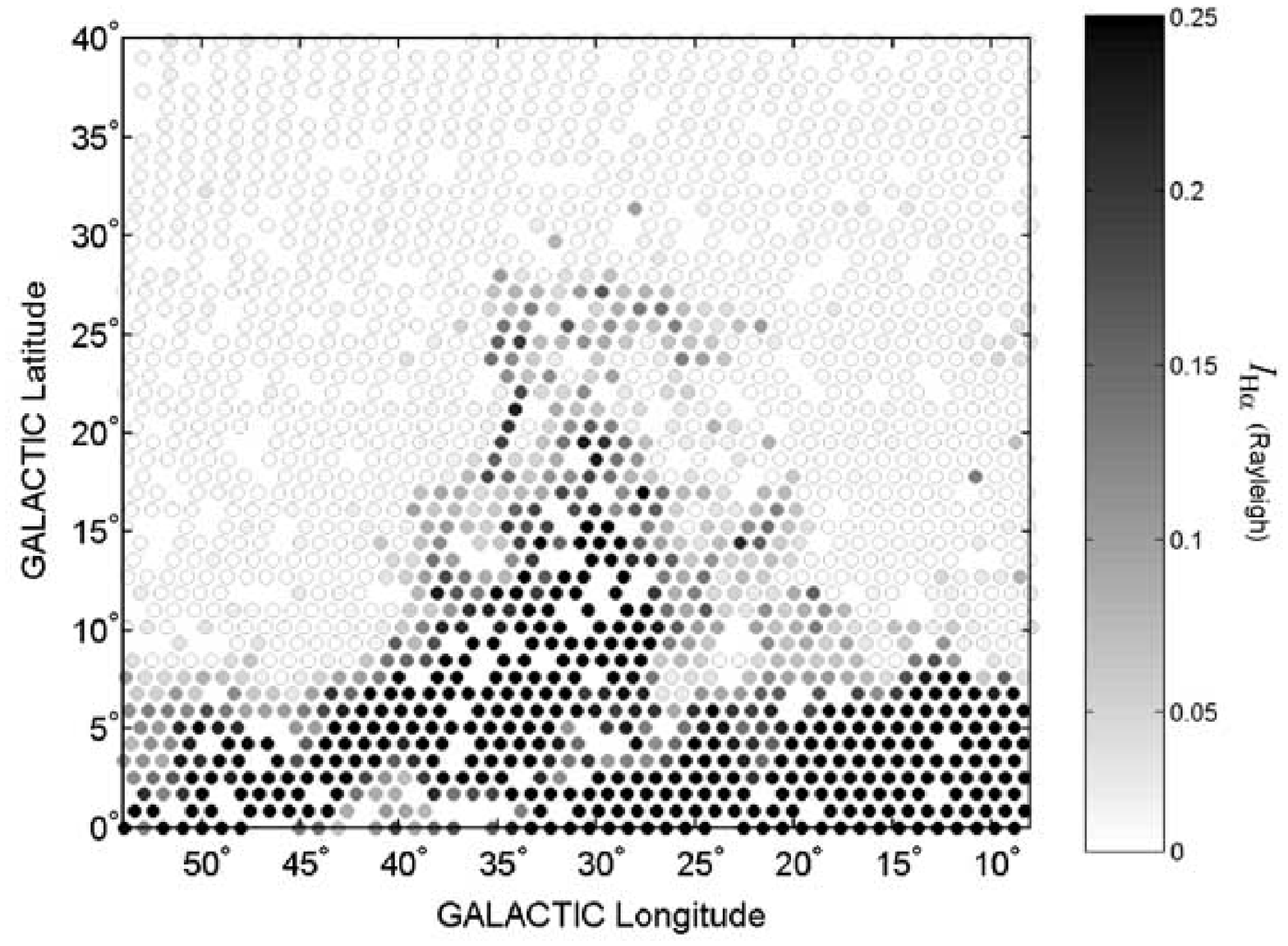} \caption{Region of the superbubble in \Ha, integrated
over 55 -- 95~\kms LSR, as observed with the Wisconsin H-Alpha
Mapper (WHAM; \citet{WHAM}). The figure conveys the nature of the
WHAM survey.  Each circle is centered at the coordinates of a WHAM
observation, but for clarity is only half the diameter of the
$1\arcdeg$ WHAM beam.  The superbubble is a major feature in \Hp
as well as \HI, but unlike the neutral hydrogen,
the ionized hydrogen fills the area and is not
concentrated at its edges.  Empty spots occur where bright
foreground stars contaminate the data.}
\end{figure*}

\clearpage
\begin{figure*}
\plotone{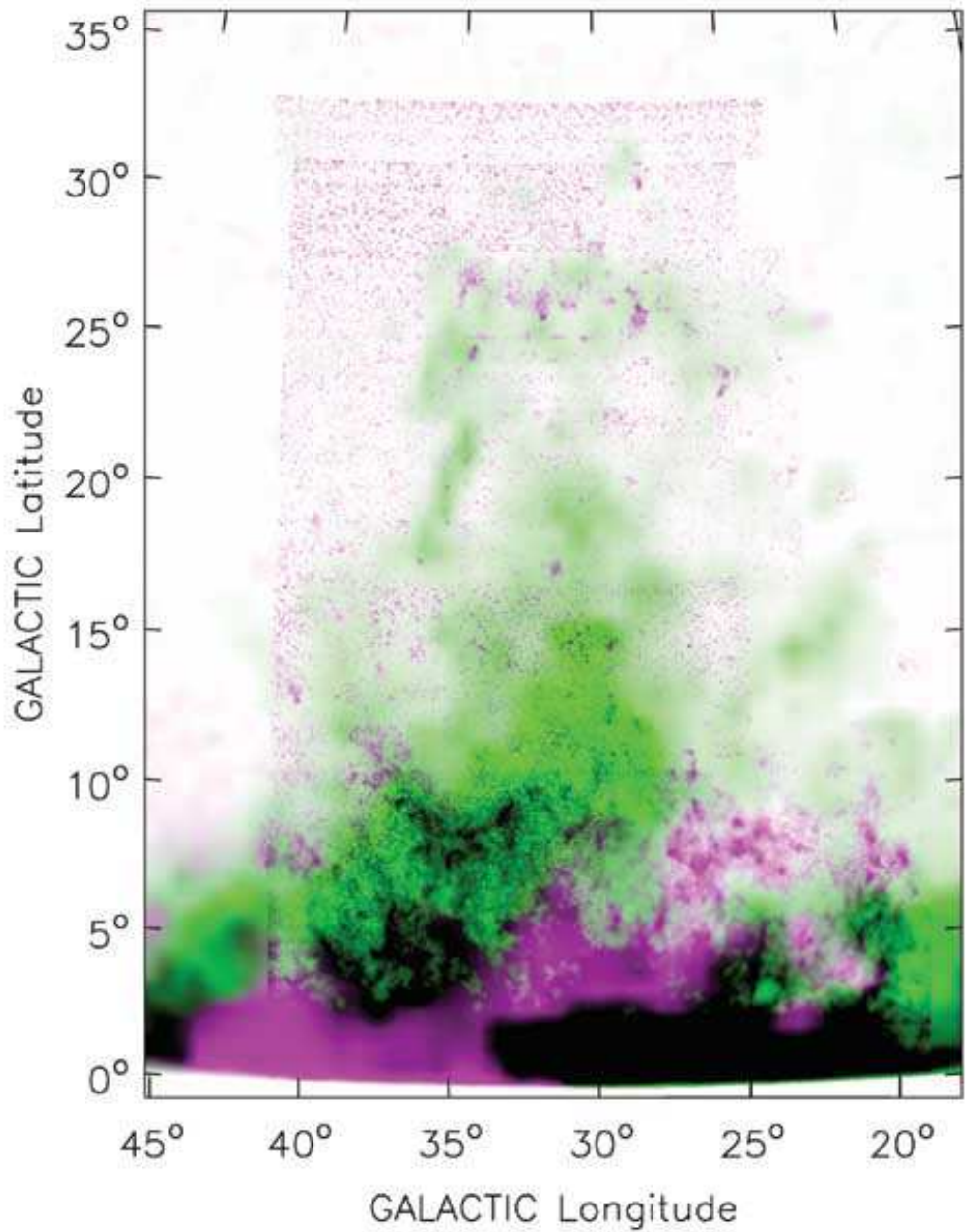} \caption{\HI data from Figure~3 (purple) and
interpolated \Ha data from Figure~6 (green). The diagonal purple
stripe at the bottom results from extinction of the \Ha by dust in
the Great Rift. There is detailed correspondence between \HI and
\Ha for
 many features, e.~g., the tips
of the \HI whiskers at $\ell, b \approx  40\arcdeg+15\arcdeg$
and $30\arcdeg+15\arcdeg$,  clouds at $29\arcdeg+31\arcdeg$
and $22\arcdeg+25\arcdeg$, and indeed, the
Plume itself. The ionized components of some of the smaller clouds
appear shifted to higher longitude than the \HI, but this
is probably  the effect of the sparsity of the \Ha survey and its
$1\arcdeg$ beam size. The \Ha emission connects the cap
continuously to the Galactic disk and thus suggests that this
 system is a singular phenomenon of gigantic proportions. }
\end{figure*}

\clearpage
\begin{figure*}
\plotone{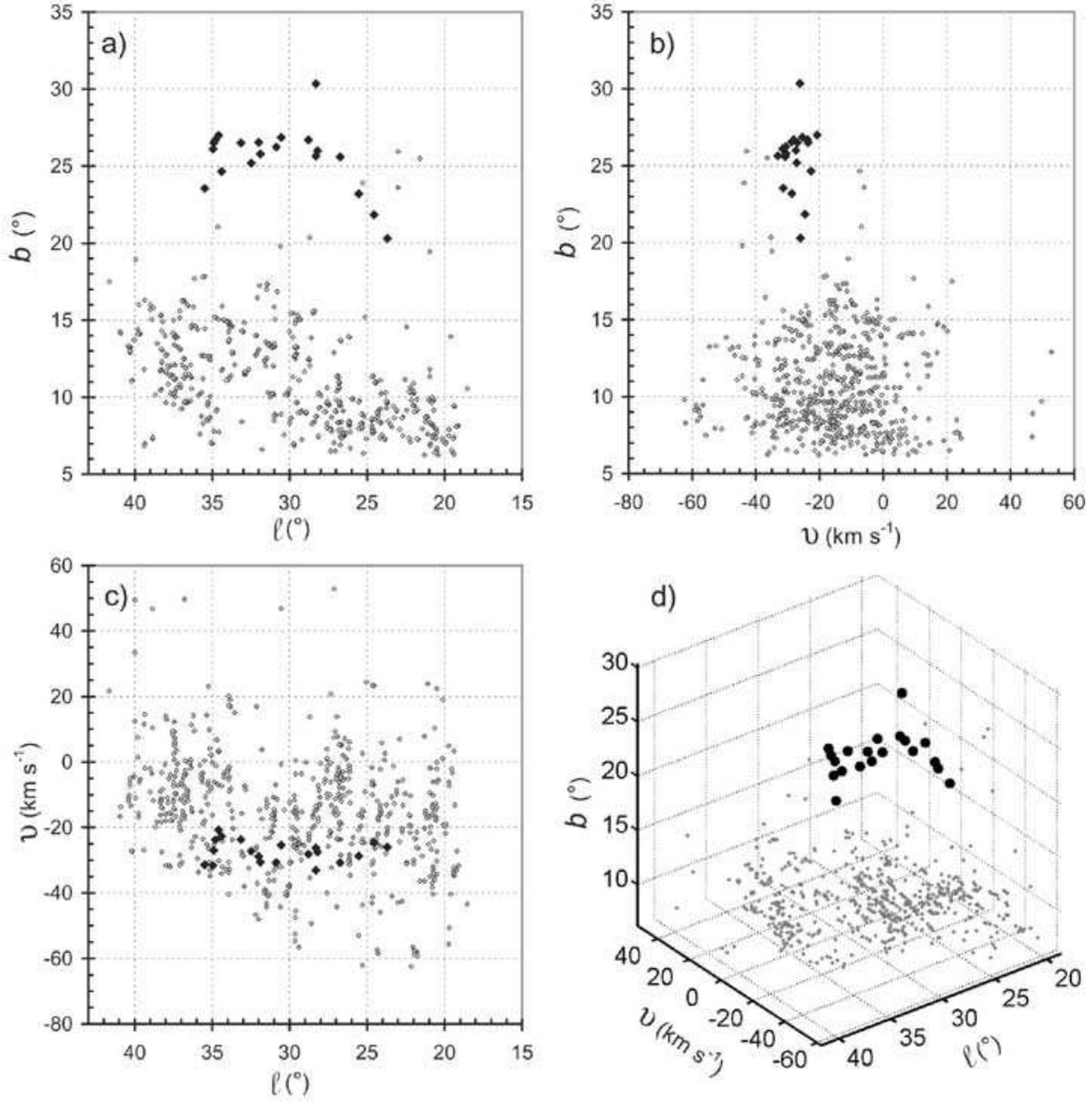} \caption{The superbubble system kinematics in $(\ell,
b,\upsilon)$ space, where $\upsilon (\ell, b, V_{\rm LSR}) \equiv
V_{\rm LSR} \sec (b)  - V_t (\ell)$ is the ``deprojected
velocity.'' Objects in circular Galactic rotation at the tangent
point
 should all have a similar
$|\upsilon| \approx 0$.  Larger filled points mark measurements on
the Plume itself, the same set shown in Figure~4. Panels a)
through c) show three 2D projections and panel d) a 3D plot of the
same data. The Plume is kinematically compact compared to other
parts of the  system and has a slight velocity offset.}
\end{figure*}

\clearpage
\begin{figure*}
\plotone{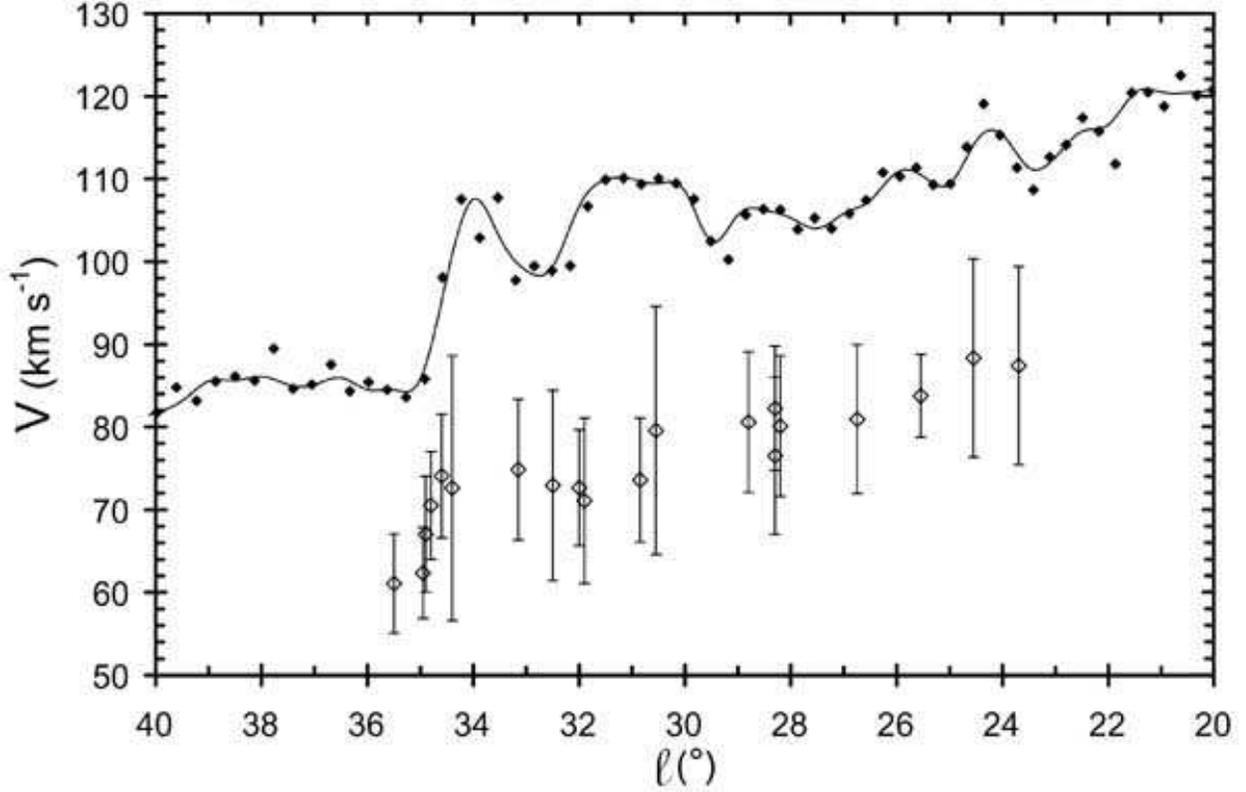} \caption{Velocity  of the Plume components compared
to that of molecular clouds at the tangent point. The filled points are measurements of
the $^{12}$CO terminal velocity in the Galactic plane
\citep{COcurve}. The curve is a cubic spline interpolation fit to
the median filtered $^{12}$CO velocities. The open symbols show
the Plume's \HI velocity,  $V_{\rm LSR}\,\sec (b)$, and the
vertical bars show the line FWHM. The Plume traces the
kinematics of the molecular gas, but with an offset velocity. }
\end{figure*}

\clearpage
\begin{figure*}
\plotone{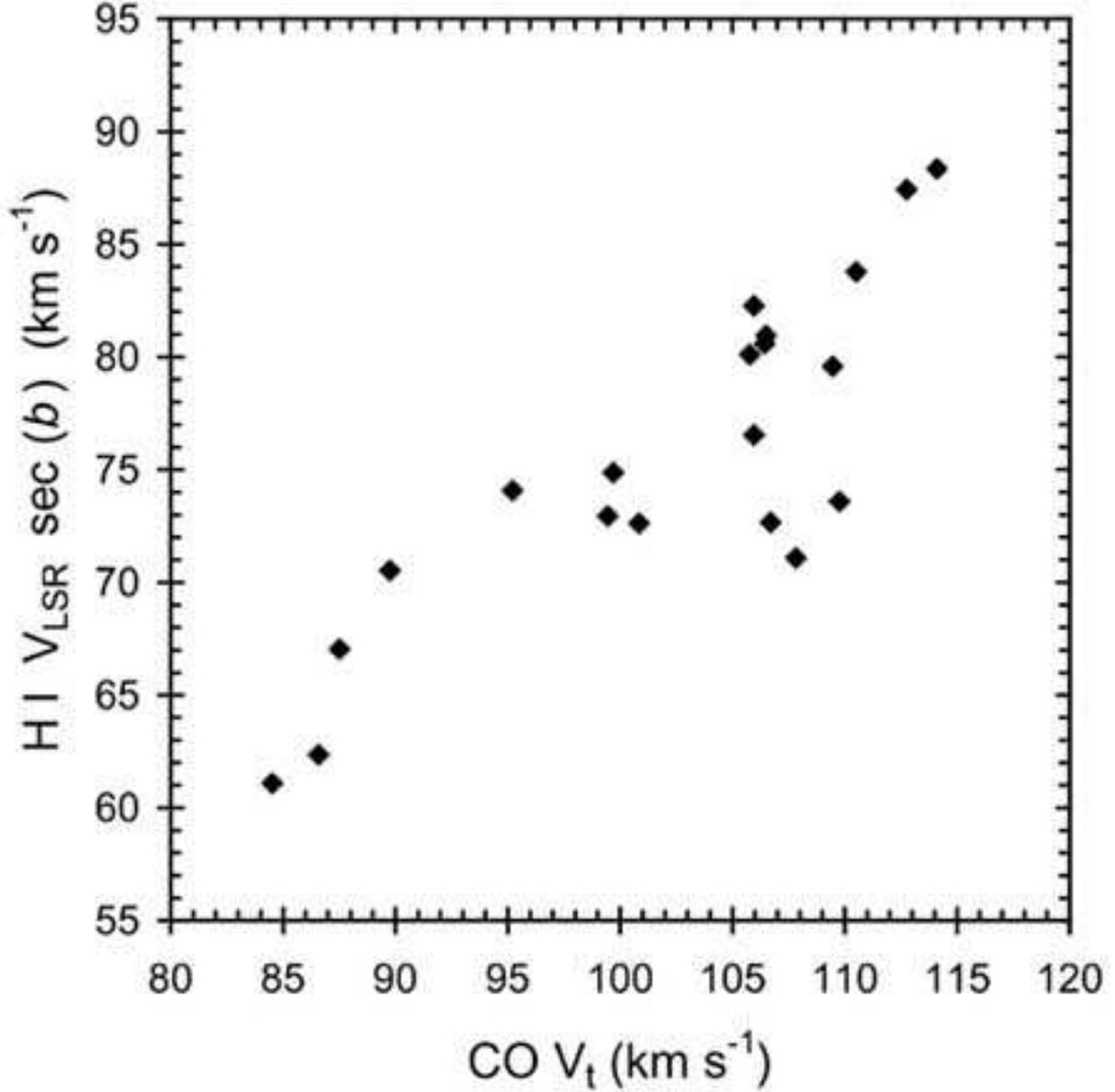} \caption{Velocities of Plume \HI features shown
in Figs.~4, 8 and 9, $V_{\rm LSR}\,\sec (b)$, plotted versus the
$^{12}$CO terminal velocity at the same longitude in the plane.
The Pearson correlation coefficient of the two sets is 85.3\%.
Despite its location at $b \approx 26\arcdeg$ ($z > 3$~kpc), the
Plume still bears  a detailed imprint of  the rotation of the
Galactic disk.}
\end{figure*}

\clearpage
\begin{figure*}
\plotone {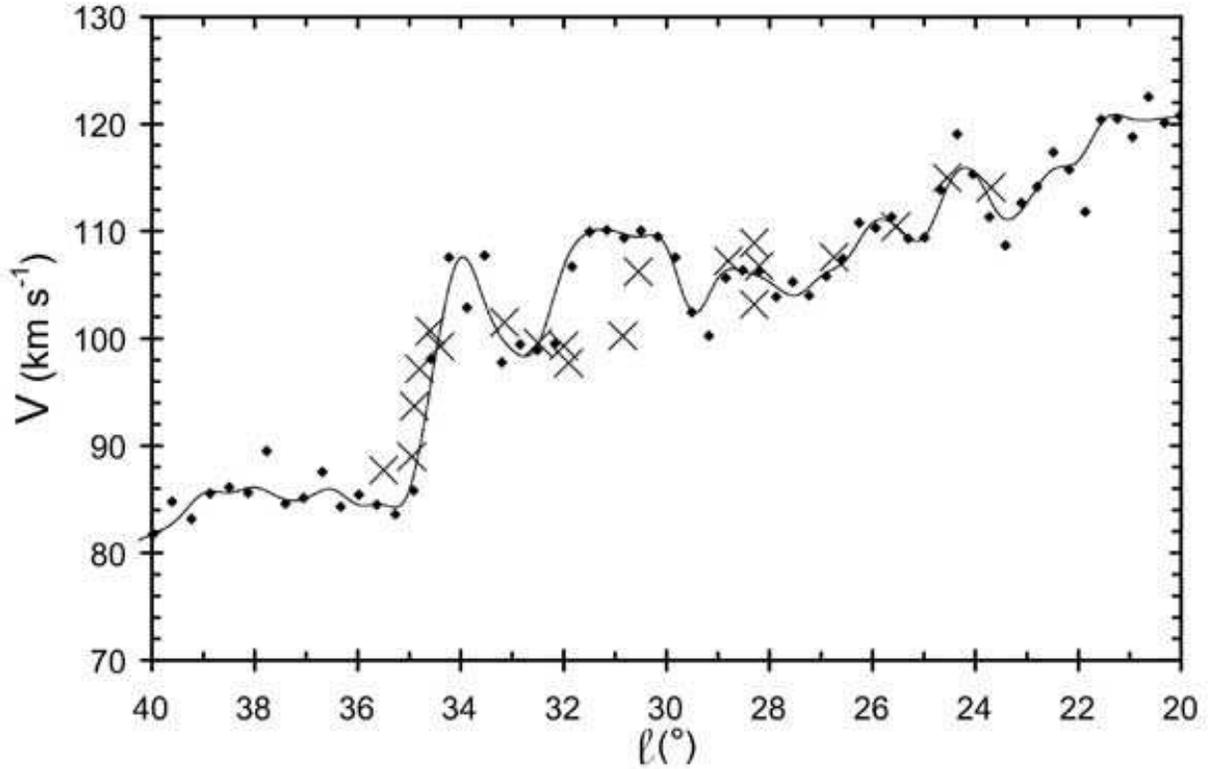} \caption{Velocity structure of the Plume,
compensated for lag, compared with that of molecular gas at the
tangent point. These are the same data as in Fig.~9, but with the \HI
velocities (marked here with crosses) increased by 26.6~\kms. The
close agreement of these measurements shows that the Plume
at $z > 3$ kpc shares the
same kinematics as molecular clouds in the Galactic plane, even in
considerable detail.  Its kinematics are thus dominated by Galactic
rotation.}
\end{figure*}

\clearpage
\begin{figure*}
\plotone {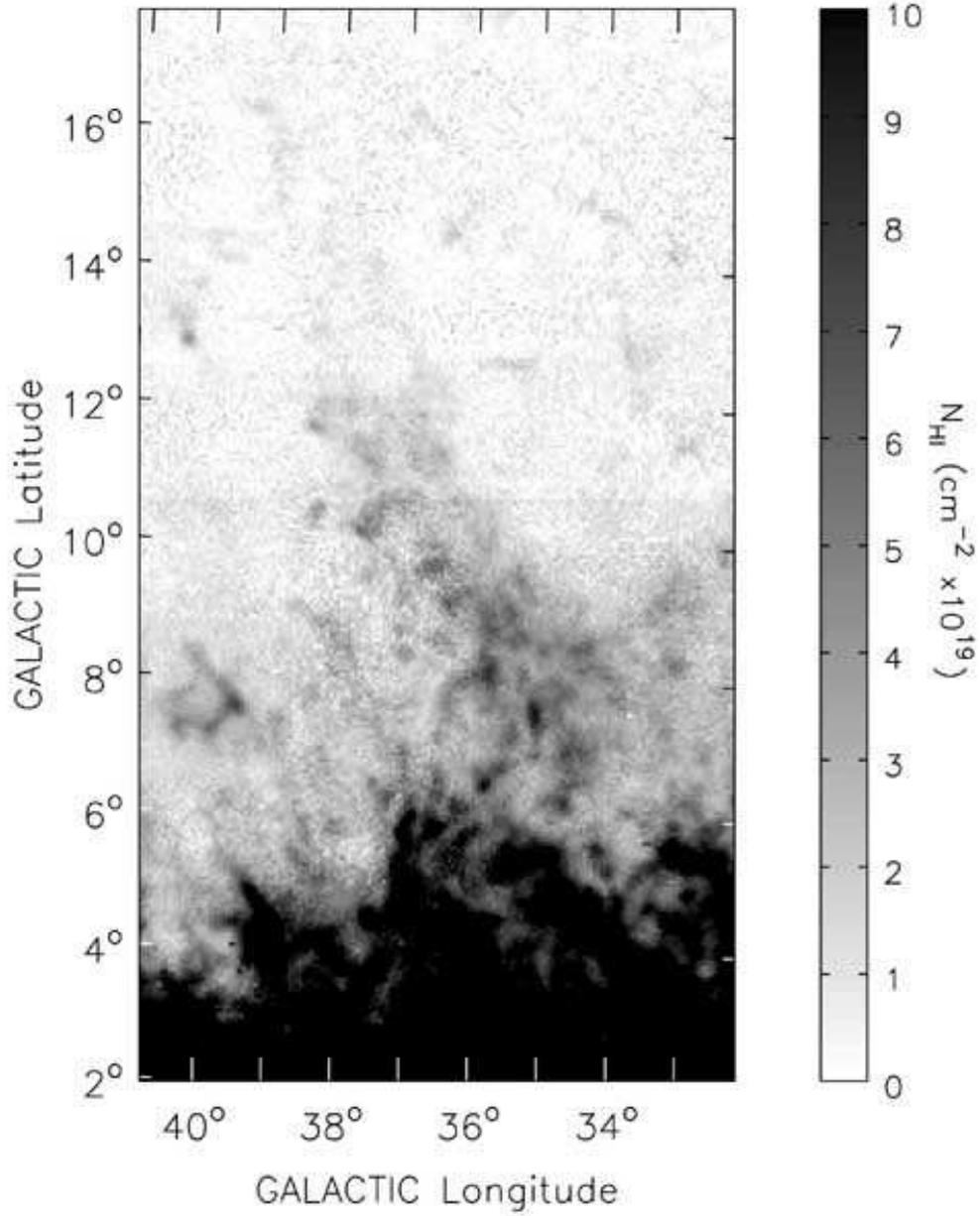} \caption{
GBT \HI observations of a prominent whisker of \HI which
may be the wall of the superbubble.  This image is integrated over
$ 70 \leq V_{\rm LSR} \leq 90$~\kms, velocities which cover the
tangent points in this direction, where $1\arcdeg$ corresponds to
120~pc. The whisker thus extends about 1.5~kpc into the halo.}
\end{figure*}

\clearpage
\begin{figure*}
\plotone{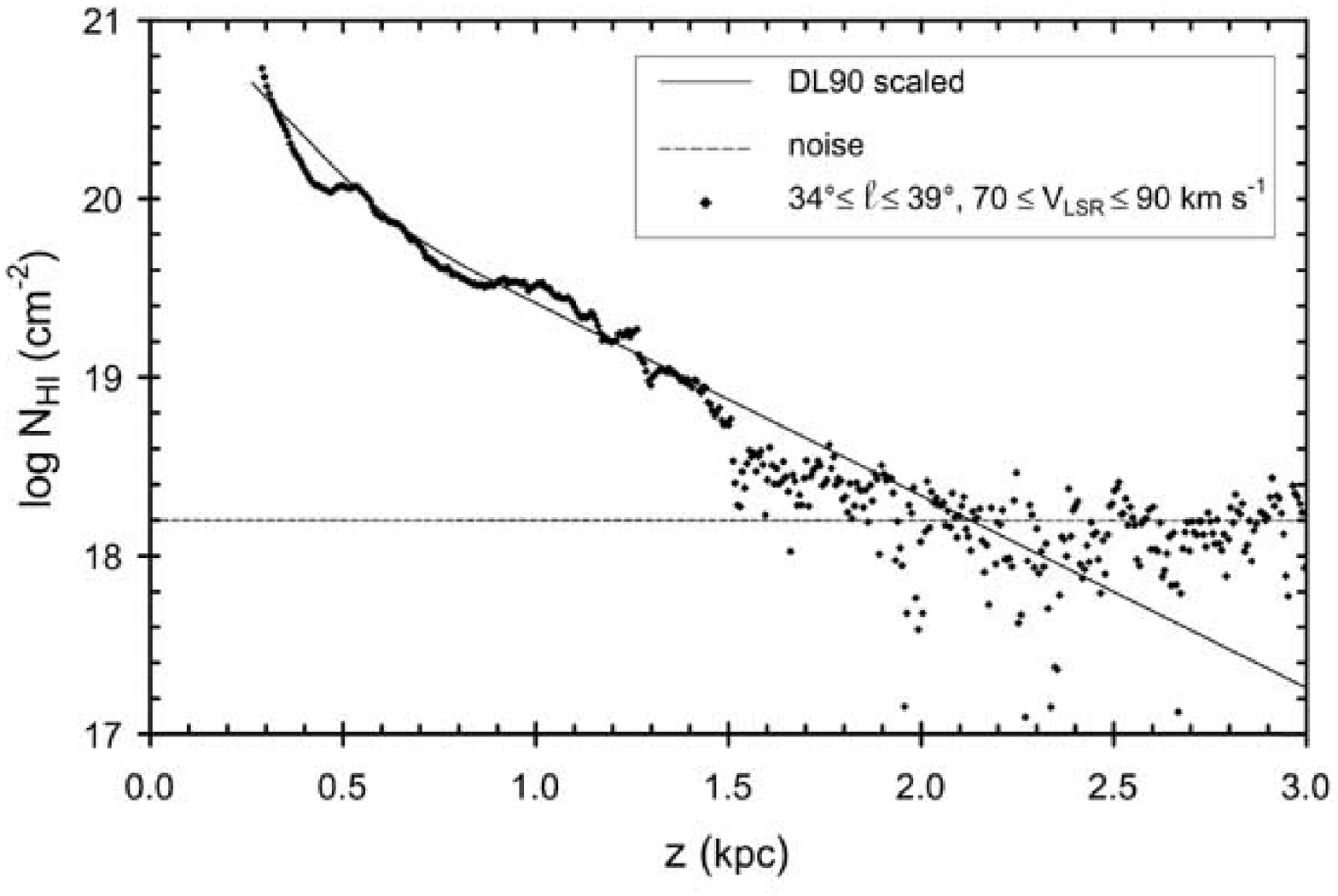} \caption{Vertical density structure of the
whisker from Fig.~12. The points show \NHI of the \HI whisker
averaged over $5\arcdeg$ of longitude, plotted against distance
from the plane assuming that the whisker is at the tangent point.
The top of the whisker is at $z \approx 1.5$~kpc. The solid curve
shows the empirical \NHI$(z)$ derived by \citet{DLHI} over the
inner Galaxy, scaled to a 1.6~kpc path length. The agreement
between the curve and points suggests that the vertical structure
of the whisker is essentially the vertical structure of the
general ISM.  The whisker is thus not likely to be material thrust
up from the plane, but rather a wall of gas swept up from the
side, as would be expected if it is the wall of a superbubble.  }
\end{figure*}

\clearpage
\begin{figure*}
\plotone{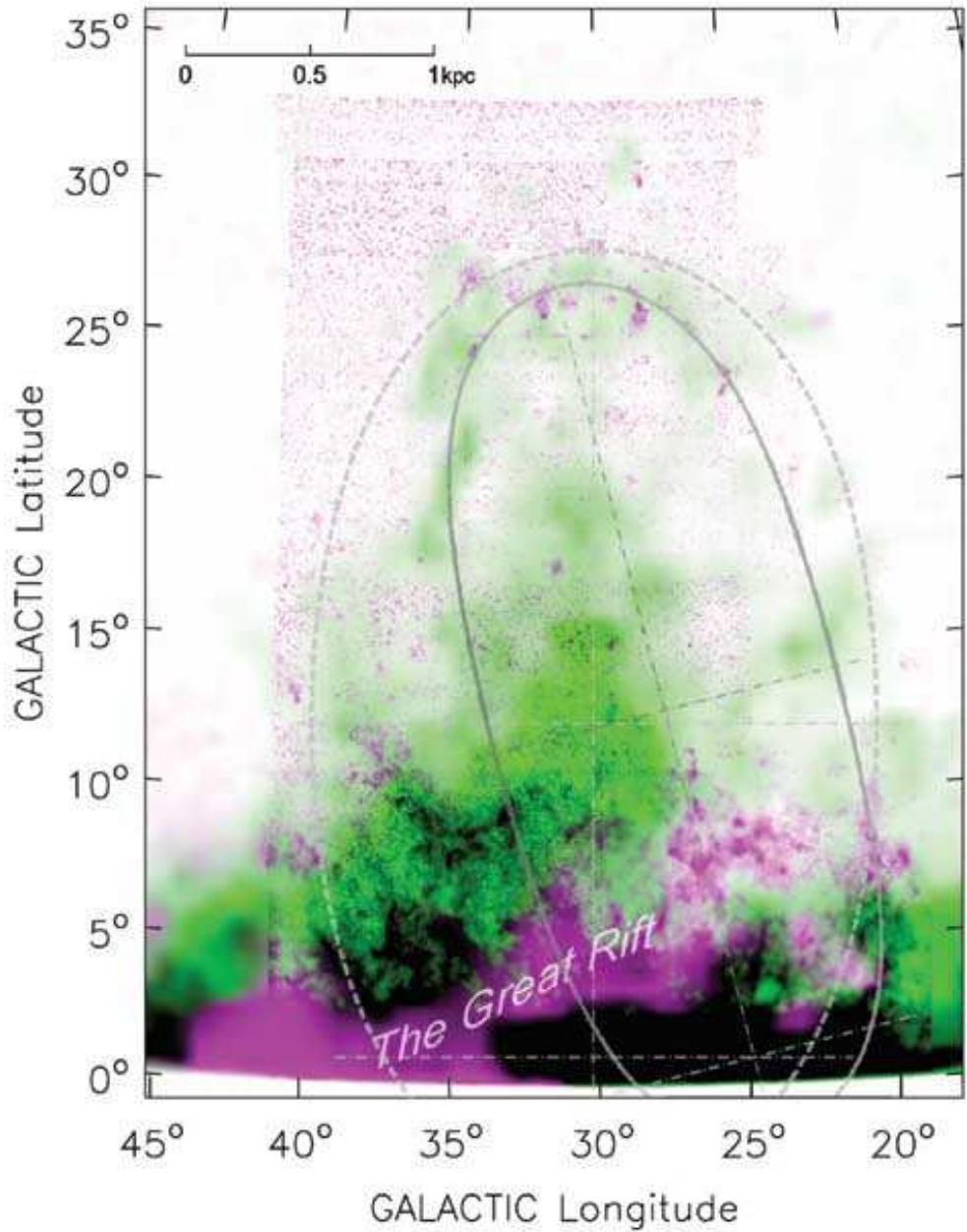} \caption{\HI and \Ha images of the Ophiuchus
superbubble (from Fig.~7) with two Kompaneets models that might
describe its structure.  They share the same latitude of origin
slightly above the plane, but differ in aspect ratio and tilt.  A
distance scale is given at the top of the Figure. The implied ages
and energetics of the two models are quite similar, and, because
the system is so large, are reasonably robust to small changes in
initial conditions. A tilt is expected from the change in
gravitational potential over so large a system, and matches the
general tilt of the \HI whiskers.}
\end{figure*}


\begin{thebibliography}{}

\bibitem[Asgekar et~al.(2005)]{Asgekar} Asgekar,
A., English, J., {Safi-{H}arb}, S.,
 \& Kothes, R. 2005, \aj, 130, 674

\bibitem[{de Avillez} \& Berry(2001)]{AvillezBerry2001}
{de Avillez}, M.~A. \&  Berry, D.~L. 2001 \mnras, 328, 708


\bibitem[{de Avillez} \& {Mac Low}(2001)]{Avillez} {de Avillez},
M.~A. \&  {Mac Low}, M.~M. 2001, \apjl, 551, L57

\bibitem[{Barnab\`{e}} et~al.(2006)]{Barnabe} {Barnab\`{e}}, M.,
 Ciotti, L., Fraternali, F., \& Sancisi, R. 2006, \aap, 446, 61

\bibitem[Basu et~al.(1999)]{KompanBubble}
 Basu, S., Johnstone, D., \& Martin, P.~G. 1999, \apj, 516, 843

\bibitem[Bingham(1967)]{Bingham} Bingham, R.~G. 1967, \mnras, 137, 157

\bibitem[{Bisnovatyj-{K}ogan} \& Silich(1995)]{Bisnovatyj}
{Bisnovatyj-{K}ogan}, G.~S. \& Silich, S.~A. 1995,
Rev.~Mod.~Phys., 67, 661

\bibitem[Bissantz et~al.(2003)]{Bissantz}
 Bissantz, N., Englmaier, P., \& Gerhard, O. 2003, \mnras, 340, 949

\bibitem[{Bland-{H}awthorn} \& Maloney(2002)]{BlandHawthornMaloney}
{Bland-{H}awthorn}, J., \&  Maloney, P.~R. 2002, in ASP
Conf.~Ser.~254, Extragalactic Gas at Low Redshift,
ed.~J.~S.~Mulchaey \& J.~Stocke (San Francisco: ASP), 267

\bibitem[Breitschwerdt \& {de Avillez}(2006)]{Breitschwerdt06}
Breitschwerdt, D. \& {de Avillez}, M.~A. 2006 \aap, 452, L1


\bibitem[Brinks \& Bajaja(1986)]{Brinks86}
Brinks, E. \& Bajaja, E.~1986, \aap, 169, 14

\bibitem[Burton \& Gordon(1978)]{Burton}
 Burton, W.~B. \& Gordon, M.~A.~1978, \aap, 63, 7

\bibitem[Callaway et~al.(2000)]{Callaway2000} Callaway, M.~B.,
 Savage, B.~D., Benjamin, R.~A., Haffner, L.~M., \& Tufte, S.~L. 2000, \apj, 532, 943

\bibitem[Castor et~al.(1975)]{SpherBubble}
 Castor, J., {McCray}, R., \& Weaver, R. 1975, \apjl, 200, L107

\bibitem[Clemens(1985)]{COcurve}
 Clemens, D.~P. 1985, \apj, 295, 422

\bibitem[Collins et~al.(2002)]{ballist}
 Collins, J.~A., Benjamin, R.~A., \& Rand, R.~J. 2002, \apj, 578, 98

\bibitem[Cooper et~al.(2004)]{Cooper04}
Cooper, R.~L., Guerrero, M.~A., Chu, Y.-H., Chen, C.-H.~R., \&
Dunne, B.~C. 2004 \apj, 605, 751

\bibitem[Dehnen \& Binney(1998)]{GalPot}
 Dehnen, W. \& Binney, J. 1998, \mnras, 294, 429

\bibitem[Dickey \& Lockman(1990)]{DLHI}
 Dickey, J.~M. \& Lockman, F.~J.~1990, \araa, 28, 215

\bibitem[Diplas \& Savage(1994)]{DiplasSavage} Diplas, A., \& Savage,
   B.~D. 1994, \apj, 427, 274

\bibitem[Dove et~al.(2000)]{DoveShullFerrara} Dove, J.~B.,
 Shull, J.~M., \& Ferrara, A. 2000, \apj, 531, 846

\bibitem[{Ehlerov\'{a}} \& {Palou\v{s}}(2005)]{Ehlerova} {Ehlerov\'{a}}, S.
\& {Palou\v{s}}, J. 2005, \aap, 437, 101

\bibitem[Elmegreen \& Lada(1977)]{ElmegreenLada77}
Elmegreen, B.~G. \&  Lada, C.~J. 1977 \apj, 214, 725

\bibitem[English et~al.(2000)]{English} English, J., Taylor, A.~R.,
Mashchenko, S.~Y., Irwin, J.~A., Basu, S., \& Johnstone, D. 2000,
\apj, 533, L25

\bibitem[Fraternali \& Binney(2006)]{Fraternali} Fraternali, F. \&
  Binney, J.~J. 2006, \mnras, 366, 449

\bibitem[Haffner et~al.(1998)]{WIM}
 Haffner, L.~M., Reynolds, R.~J., \& Tufte, S.~L. 1998, \apjl, 501, L83

\bibitem[Haffner et~al.(2003)]{WHAM}
 Haffner, L.~M., Reynolds, R.~J., Tufte, S.~L., Madsen, G.~J., Jaehnig, K.~P.,
 \& Percival, J.~W. 2003, \apjs, 149, 405

\bibitem[Hartmann \& Burton(1997)]{LD}
 Hartmann, D. \& Burton, W.~B.~1997, Atlas of Galactic Neutral Hydrogen,
 (Cambridge, UK: Cambridge University Press)

\bibitem[Haslam et~al.(1982)]{Haslam}
 Haslam, C.~G.~T., Salter, C.~J., Stoffel, H., \& Wilson, W.~E. 1982, \aaps, 47, 1

\bibitem[Heiles(1979)]{Heiles79} Heiles, C. 1979, \apj, 229, 533

\bibitem[Heiles(1984)]{Heiles84} Heiles, C. 1984, \apjs, 55, 585

\bibitem[Howk \& Savage(2000)]{Howk} Howk, J.~C., \& Savage, B.~D. 2000,
 \aj, 119, 644

\bibitem[Hu(1981)]{Hu81} Hu, E.~M. 1981, \apj, 248, 119

\bibitem[Igumenshchev et~al.(1990)]{Igumen90}
Igumenshchev, I.~V., Shustov, B.~M., \&  Tutukov, A.~V. 1990 \aap,
234, 396

\bibitem[Kim et~al.(1999)]{Kim} Kim, S., Dopita, M.~A., {Staveley-{S}mith}, L.,
 \& Bessell, M.~S.~1999, \aj, 118, 2797

\bibitem[Kompaneets(1960)]{Kompan60} Kompaneets, A.~S. 1960, Soviet Phys.~Dokl., 5, 46

\bibitem[Koo et~al.(1992)]{Koo} Koo, B.-C., Heiles, C., \&
  Reach, W.~T. 1992, \apj.~390, 108

\bibitem[Koo \& {McKee}(1990)]{KooMcKee0} Koo, B.-C.~\& {McKee}, C.~F. 1990, \apj, 354, 513

\bibitem[Koo \& {McKee}(1992a)]{KooMcKee1} Koo, B.-C.~\& {McKee}, C.~F. 1992, \apj, 388, 93

\bibitem[Koo \& {McKee}(1992b)]{KooMcKee2} Koo, B.-C.~\& {McKee}, C.~F. 1992, \apj, 388, 103

\bibitem[Korpi et~al.(1999)]{Korpi99} Korpi, M.~J.,
Brandenburg, A., Shukurov, A., \& Tuominen, I. 1999, \aap, 350,
230

\bibitem[Kudoh \& Basu(2004)]{KudohBasu2004} Kudoh, T. \& Basu, S. 2004, \aap, 423, 183

\bibitem[Lockman(1984)]{FJL84}
 Lockman, F.~J. 1984, \apj, 283, 90

\bibitem[Lockman et~al.(1986)]{FJL86}
 Lockman, F.~J., Jahoda, K., \& {McCammon}, D. 1986, \apj, 302, 432

\bibitem[Lockman(1989)]{FJL89}
 Lockman, F.~J. 1989, \apjs, 71, 469

\bibitem[Lockman(2002a)]{FJL02}
 Lockman, F.~J. 2002, \apjl, 580, L47

\bibitem[Lockman(2002b)]{FJL02stray}
 Lockman, F.~J. 2002, in ASP Conf.~Ser.~278, Single-Dish Radio Astronomy:
Techniques and Applications, ed.~S.~Stanimirovic et~al. (San
Francisco: ASP), 397

\bibitem[Lockman \& Pidopryhora(2005)]{LP04}
 Lockman, F.~J. \& Pidopryhora, Y. 2005, in ASP Conf.~Ser.~331, Extra-planar
 Gas, ed.~R.~Braun (San Francisco: ASP), 59

\bibitem[Lockman \& Condon(2005)]{FJLstray}
 Lockman, F.~J. \& Condon, J.~J. 2005, \aj, 129, 1968

\bibitem[Maciejewski et~al.(1996)]{Maciejewski96} Maciejewski, W., Murphy, E.~M.,
Lockman, F.~J., \& Savage, B.~D. 1996, \apj, 469, 238

\bibitem[{Mac Low} \& {McCray}(1988)]{MacLow}
 {Mac Low}, M.~M. \& {McCray}, R. 1988, \apj, 324, 776

\bibitem[{Mac Low} et~al.(1989)]{MacLow89} {Mac Low}, M.~M., {McCray}, R., \& Norman, M.~L. 1989, \apj, 337, 141

\bibitem[Madsen \& Reynolds(2005)]{Madsen2005}
 Madsen, G.~J. \& Reynolds, R.~J. 2005, \apj, 630, 925

\bibitem[Martins et~al.(2005)]{Ostars}
 Martins, F., Schaerer, D., \& Hillier, D.~J. 2005, \aap, 436, 1049

\bibitem[Matsushita et~al.(2005)]{Matsushita}
Matsushita, S., Kawabe, R., Kohno, K., Matsumoto, H., Tsuru,
T.~G., \& {Vila-{V}ilar\'{o}}, B. 2005, \apj, 618, 712

\bibitem[{McClure-{G}riffiths} et~al.(2000)]{McClure2000}
{McClure-{G}riffiths}, N.~M., Dickey, J.~M., Gaensler, B.~M.,
Green, A.~J., Haynes, R.~F., \& Wieringa, M.~H. 2000, \aj, 119,
2828

\bibitem[{McClure-{G}riffiths} et~al.(2002)]{McClure-Griffiths}
{McClure-{G}riffiths}, N.~M., Dickey, J.~N., Gaensler, B.~M., \&
Green, A.~J. 2002, \apj, 578, 176

\bibitem[{McCray} \& Kafatos(1987)]{McCrayKafatos87}
{McCray}, R. \& Kafatos, M. 1987, \apj, 317, 190

\bibitem[Motte et~al.(2003)]{Motte}
Motte, F., Schilke, P., \& Lis, D.~C. 2003, \apj, 582, 277

\bibitem[Oey \& Clarke(1997)]{OeyClarke} Oey, M.~S. \& Clarke, C.~J. 1997, \mnras, 289, 570

\bibitem[Oey (2004)]{Oey2004} Oey, M.~S. 2004, \apss, 289, 269

\bibitem[Oey \& {Garc\'{i}a-{S}egura}(2004)]{OeyGarcia} Oey, M.~S. \& {Garc\'{i}a-{S}egura}, G. 2004, \apj, 613, 302

\bibitem[{Pikel'ner} \& {Shcheglov}(1969)]{PikelnerShcheglov69}
{Pikel'ner}, S.~B. \&  Shcheglov, P.~V. 1969, Soviet Astronomy,
12, 757

\bibitem[Putman et~al.(2003)]{Putman} Putman, M.~E., {Bland-{H}awthorn}, J., Veilleux, S.,
 Gibson, B.~K., Freeman, K.~C., \& Maloney, P.~R. 2003, \apj, 597, 948

\bibitem[Reynolds(1985)]{ReynoldsT8000}
 Reynolds, R.~J. 1985, \apj, 294, 256

\bibitem[Sancisi(1999)]{Sancisi} Sancisi, R. 1999, \apss, 269, 59

\bibitem[Savage et~al.(1990)]{Savage90} Savage, B.~D., Massa,
  D., \& Sembach, K. 1990, \apj, 355, 114

\bibitem[Savage et~al.(1997)]{SavageSemLu} Savage, B.~D.,
 Sembach, K.~R., \& Lu, L. 1997, \aj, 113, 2158

\bibitem[Savage et~al.(2001)]{Savage2001} Savage, B.~D., Sembach, K.~R., \& Howk, J.~C. 2001, \apj, 547, 907

\bibitem[Smith et~al.(1978)]{Smith}
 Smith, L.~F., Biermann, P., \& Mezger, P.~G. 1978, \aap, 66, 65

\bibitem[Smith(2006)]{nsmith06} Smith, N. 2006, \mnras, 367, 763

\bibitem[Snowden et~al.(1997)]{ROSAT}
 Snowden, S.~L., Egger, R., Freyberg, M.~J., Plucinsky, P.~P., Schmitt, J.~H.~M.~M.,
{Tr\"{u}mper}, J., Voges, W., {McCammon}, D., \& Sanders, W.~T.
1997, \apj, 485, 125

\bibitem[{Stanimirovic} et~al.(1999)]{Stanimirovic99}
Stanimirovic, S.,  {Staveley-{S}mith}, L., Dickey, J.~M., Sault,
R.~J., \& Snowden, S.~L. 1999, \mnras, 302, 417

\bibitem[{Staveley-{S}mith} et~al.(1997)]{Staveley97}
{Staveley-{S}mith}, L., Sault, R.~J., Hatzidimitriou, D.,
Kesteven, M.~J., \& McConnell, D. 1997, \mnras, 289, 225

\bibitem[{Tenorio-{T}agle} et~al.(1987)]{T-TBR}
{Tenorio-{T}agle}, G., Bodenheimer, P., \& Rozyczka, M. 1987 \aap,
182, 120

\bibitem[{Tenorio-{T}agle} \& Bodenheimer(1988)]{TenorioTagleARAA88}
{Tenorio-{T}agle}, G. \& Bodenheimer, P. 1988 \araa, 26, 145

\bibitem[{Tenorio-{T}agle} et~al.(1990)]{Tenorio90}
{Tenorio-{T}agle}, G., Rozyczka, M., \& Bodenheimer, P. 1990 \aap,
237, 207

\bibitem[Tomisaka \& Ikeuchi(1986)]{TomisakaIkeuchi86}
Tomisaka, K. \& Ikeuchi, S. 1986, \pasj, 38, 697

\bibitem[Tomisaka(1998)]{Tomisaka98}
Tomisaka, K. 1998 \mnras, 298, 797

\bibitem[Tripp et~al.(2003)]{Tripp} Tripp, T.~M., Wakker, B.~P., Jenkins, E.~B.,
Bower, C.~W., Danks, A.~C., Green, R.~F., Heap, S.~R., Joseph,
C.~L., Kaiser, M.~E., Linsky, J.~L., \& Woodgate, B.~E. 2003, \aj,
125, 3122

\bibitem[Tufte et~al.(1998)]{Tufteetal98} Tufte, S.~L.,
Reynolds, R.~J., \& Haffner, L.~M. 1998, \apj, 504, 773

\bibitem[Tufte et~al.(2002)]{Tufteetal02} Tufte, S.~L., Wilson, J.~D.,
Madsen, G.~J., Haffner, L.~M., \& Reynolds, R.~J. 2002, \apj, 572,
L153

\bibitem[Veilleux et~al.(2005)]{GalWinds}
 Veilleux, S., Cecil, G., \& {Bland-{H}awthorn}, J. 2005, \araa, 43, 769

\bibitem[Voit(1988)]{Voit88} Voit, G.~M. 1988, \apj, 331, 343

\bibitem[Walter \& Brinks(1999)]{WalterBrinks} Walter, F. \& Brinks, E. 1999, \aj 118, 273

\bibitem[Weaver et~al.(1977)]{Weaver}
Weaver, R., {McCray}, R., Castor, J., Shapiro, P., \& Moore, R.
1977, \apj, 218, 377

\bibitem[{Wei\ss} et~al.(1999)]{Weiss} {Wei\ss}, A., Walter, F., Neininger, N, \& Klein,
U. 1999, \aap, 345, L23

\bibitem[Williams(1973)]{Williams} Williams, D.~R.~W. 1973, \aaps, 8, 505

\bibitem[Willingale et~al.(2003)]{Willingale}
Willingale, R., Hands, A.~D.~P., Warwick, R.~S., Snowden, S.~L.,
\& Burrows, D.~N. 2003, \mnras, 343, 995

\bibitem[Wolfire et~al.(1995)]{WolfirePot}
 Wolfire, M.~G., {McKee}, C.~F., Hollenbach, D., \& Tielens, A.~G.~G.~M. 1995, \apj, 453, 673

\end{thebibliography}
\end{document}